\documentclass{aa}
\usepackage{txfonts}
\usepackage[T1]{fontenc}

\usepackage{microtype,siunitx,booktabs}
\usepackage{amsmath}
\usepackage{mathtools, stmaryrd}
\usepackage{xparse}
\usepackage{diagbox}
\usepackage{adjustbox}
\usepackage[normalem]{ulem}
\begin{document}

\title{Photometric determination of rotation axis inclination, rotation rate, and mass of rapidly rotating intermediate-mass stars\thanks{The whole set of synthetic spectra and their associated data are available at the CDS via anonymous ftp to cdsarc.cds.unistra.fr (130.79.128.5) or via https://cdsarc.cds.unistra.fr/cgi-bin/qcat?J/A+A/}}

\author{Axel Lazzarotto\inst{1}, Alain Hui-Bon-Hoa\inst{1} and Michel Rieutord\inst{1}}
\institute{
IRAP, Universit\'e de Toulouse, CNRS, UPS, CNES,
14, avenue \'{E}douard Belin, F-31400 Toulouse, France
\\
\email{Axel.Lazzarotto, Alain.Hui-Bon-Hoa, Michel.Rieutord[@irap.omp.eu]}
}

\titlerunning{Rotation axis inclination, rotation rate, and mass of early-type stars}
\authorrunning{Lazzarotto et al.}

\date{Received 12 April 2023; accepted 1 June 2023}
\abstract
{
Intermediate-mass stars are often fast rotators, and hence are centrifugally flattened and notably affected by gravity darkening. To analyse  this kind of stars properly, one must resort to 2D models to compute the visible radiative flux and to take the geometrical effect of the star inclination into account.}
{
Assuming a given stellar age and chemical composition, our aim is to derive the mass and rotation rates of main sequence fast rotating stars, along with their inclination, from photometric quantities influenced by gravity darkening.}
{
We chose three observables that vary with mass, rotation, and inclination: the temperature derived by the infrared flux method $T_\mathrm{IRFM}$, the Str\"omgren $c_1$ index, and a second index $c_2$ built in the same way as the $c_1$ index, but sensitive to the UV side of the Balmer jump. These observables are computed from synthetic spectra produced with the PHOENIX code and rely on a 2D stellar structure from the ESTER code. These quantities are computed for a grid of models in the range 2 to 7~$M_\odot$, and rotation rates from 30\% to 80\% of the critical rate. Then, for any triplet ($T_\mathrm{IRFM}$, $c_1$, $c_2$), we try to retrieve the mass, rotation rate, and inclination using a Levenberg-Marquardt scheme, after a selection step to find the most suitable starting models.}
{
Hare-and-hound tests showed that our algorithm can recover the mass, rotation rate, and inclination with a good accuracy. The difference between input and retrieved parameters is negligible for models lying on the grid and is less than a few percent otherwise. An application to the real case of Vega showed that the $u$ filter is located in a spectral region where the modelled and observed spectra are discrepant, and led us to define a new filter. Using this new filter and subsequent index, the Vega parameters are also retrieved with satisfactory accuracy.}
{
This work opens the possibility to determine the fundamental parameters of rapidly rotating early-type stars from photometric space observations.}



\maketitle

\section{Introduction}

The influence of rotation on the evolution of stars is a long-standing question for their modelling \citep{maeder09}. The major consequence of rotation is the so-called rotational mixing. It affects radiative regions thanks to the forcing of baroclinic flows and their associated instabilities, which generate a small-scale turbulence \cite[e.g.][]{mathis+18}. As a result, the lifetime on the main sequence is enhanced and surface abundances are modified \citep{maeder+14}. In addition, the interpretation of observations is also affected, in particular for fast rotating early-type stars, which is our  focus in this article.

According to \cite{ZR12}, fast rotation (equatorial velocity higher than 100~km/s) involves 50\% of early-type stars. For such stars, 1D (spherically symmetric) models may fail to correctly interpret the observations. To illustrate this point, the example of Altair is emblematic. Altair is an A7V-type star in the solar neighbourhood, 5.14~pc away from the Sun, and whose $V \sin i$ is around 240~km/s. Because of its proximity, Altair has been observed in many ways; it has been a privileged target for interferometry \cite[][]{vanbelle+01,domiciano+05,petersonetal06a,monnier+07,bouchaud+20,spalding+22}, seismology \cite[][]{buzasietal05,suarez+05,ledizes+21}, and spectroscopy \cite[][]{reiners+04b}. A long-standing question about Altair is its evolutionary stage, in short its age. Estimates using 1D models were rather inconclusive: \cite{suarez+05}  gave a range of 225-775~Myr, while \cite{domiciano+05} gave another range between 1.2 and 1.4 Gyr. However, \cite{petersonetal06a} interpreted their interferometric data with a simple Roche 2D model, and deduced from the measured polar radius and luminosity that Altair was barely off the zero-age main sequence (ZAMS). Some years later, with the advent of self-consistent 2D models\footnote{In the first 2D models that were designed, differential rotation was imposed by an ad hoc parameter \cite[e.g.][]{JMS04}. The ESTER model of \cite{ELR13} computes the differential rotation and associated meridional circulation consistently with the baroclinic distribution of pressure and temperature of the 2D stellar model.} \citep{ELR13,RELP16}, \cite{bouchaud+20} reanalysed available data of Altair from interferometry, spectroscopy, and seismology, and derived a 2D concordance model that points towards 100~Myr for Altair's age, thus indeed barely off the ZAMS.

The foregoing example of Altair is rather extreme since its equatorial velocity exceeds 300~km/s \citep{bouchaud+20}. However, it emphasises the fact that 2D models are necessary to analyse fast rotators. Altair's example interestingly shows that interferometry is well suited to determine the inclination of the rotation axis on the line of sight. This angle, $i$, is a crucial parameter since it conditions the determination of the true luminosity, the polar and equatorial radii, and the effective temperature distribution over latitude. Fast rotating stars are singled out by   gravity darkening that makes their polar regions brighter than their equatorial regions \cite[e.g.][]{R16a}. Hence, inclination strongly influences the way we see a fast rotating star, and therefore how we interpret its observations (e.g. line profiles, spectral energy distribution). For fast rotating early-type stars, the best way to determine $i$ seems to be interferometry, but the angular diameter should not be too small. Presently, 1.5 milliarcsec (mas) seems to be the minimum size, but in the near future with the forthcoming  Stellar Parameters and Images with a Cophased Array (SPICA)  to be implemented on the Center for High Angular Resolution Astronomy (CHARA) array \citep{mourard+18}, this limit may be reduced to 0.8~mas. Nevertheless, this new limit is still stringent on the number of early-type stars that can be analysed since they should typically be closer than 30~pc from the Sun. If we wish to go farther, we can only rely on spectroscopy and/or photometry, and try to get the relevant information through asteroseismic and line profile analyses, for instance.

Since we are interested in fast rotation, both spectroscopy and seismology face difficulties. The eigenspectrum of a fast rotator is very complex and requires 2D calculations of eigenmodes over a 2D model \cite[e.g.][]{reese+21}. Until now, determination of inclinations solely from asteroseismic data has been attempted only for late-type stars where the amplitude of stochastically excited waves can reasonably be guessed \citep{gizon_solanki03,gehan+21}. For early-type fast rotators where mode excitation usually comes from the $\kappa$-mechanism, no such determination has been achieved. However, if eigenmodes can be identified, then the rotation period can be deduced. With this quantity and the projected equatorial velocity $V \sin i$, obtained with a spectroscopic analysis, inclination can be estimated as   in \cite{moravveji+16}, among others,  for the slowly pulsating B-type star (SPB) star KIC7760680. Actually, since the rotation of this star is not too fast (72 km/s), 1D models are still suitable to perform its seismic analysis.

Very few attempts exist from a purely  photometric perspective,  while several spectroscopic studies have been done. For instance, \cite{fremat+05} considered rapidly rotating B-type stars and showed that neglecting gravitational darkening produces systematic errors on surface parameters, such as abundance, effective temperature $T_\mathrm{eff}$, and $V\sin i$,  emphasising the result of \cite{townsend+04} on the underestimation of $V\sin i$ in Be stars. More specifically, \cite{reiners+04b} attempted to determine Altair's inclination from a line profile analysis and found $i>68^\circ$ on a 1$\sigma$ level and above $45^\circ$ on a 2$\sigma$ level. Interferometry provides a value of  $50.7^\circ\pm1.2^\circ$ \citep{bouchaud+20} or $57.2^\circ\pm1.9^\circ$ \citep{monnier+07}. The spectroscopic exercise was applied to Vega by \cite{takeda_etal08}, who find a value of $7.2^\circ$, actually quite close  to the value provided by the concordance model of \cite{monnier_etal12}, namely $i=6.2^\circ\pm0.4^\circ$.

In the present work we propose another route to the determination of fundamental parameters of a fast rotating star. The novelty is to use the recent 2D-ESTER models in combination with the PHOENIX atmosphere models to derive a grid of spectral quantities that can be inverted to retrieve the actual physical parameters of the star. We still restrict ourselves to early-type stars of a certain age, and focus on two fundamental parameters of a fast rotating star, namely its mass and flattening, along with the inclination of the rotation axis on the line of sight. For this we compute synthetic spectra and determine three spectral quantities:  $T_{\rm IRFM}$, the temperature determined by the infrared flux method (IRFM) \citep{blackwell+80}; $c_1$,   the Str\"omgren photometric index; and $c_2$, a new photometric index using the UV part of the spectrum. These three quantities characterise the shape of the spectrum and we   show here that it is possible to use them to determine the stellar mass, the centrifugal flattening, and the inclination on the line of sight for intermediate-mass stars with fast enough rotation.

In   Sect. 2 we briefly present the numerical tools we used to generate spectra of fast rotating stellar models, then in Sect.~\ref{section:influence} we describe the properties of the observable quantities we used. Our inversion algorithm is detailed in Sect.~\ref{section:determination}, and we show how it performs using a hare-and-hounds exercise. Section~\ref{section:vega} shows the application of our method to the concrete case of Vega and how we overcome the problems that arise when dealing with this real star. In Sect.~\ref{section:discussion} we discuss questions raised by our method and by the next steps towards a full determination of fundamental parameters of early-type fast rotating stars. Our conclusions follow in Sect. 7.

\section{Computation of the observed flux of fast rotating stars}
\label{sec:computations}
For the kind of stars we are interested in, the geometry departs enough from the spherical case so that a single pair of ($T_\mathrm{eff}, \log g$) does not make sense anymore to characterise the atmosphere. The emergent flux as seen by an observer consists of the sum of the contributions of all the visible parts of the star, each with its own surface effective temperature and gravity. Therefore, for the very same star, the observed flux depends on the inclination of the rotation axis with respect to the line of sight. We thus need to calculate the spectrum as seen from an observer before computing the fluxes in various photometric bands in order to infer photometric indices. For this we use the 2D structure of the star, which provides the local effective temperature and gravity of each surface element of the visible part. These two quantities are used as inputs for the computation of a model atmosphere, which in turn is used for the synthesis of the spectrum emerging from this surface element.

\subsection{Two-dimensional internal structure code ESTER}
\label{subsec:ESTER}

The 2D internal structure is computed with  Evolution STEllaire en Rotation (ESTER), a stellar evolution code dedicated to the modelling of rapidly rotating stars \citep{REL09,REL13,ELR13,REL13c,RELP16}. ESTER relies on the following assumptions: a steady state structure that includes the baroclinic flows in radiative regions (differential rotation and meridional circulation), and hydrogen burning in the convective core. Evolution is presently only possible along the main sequence. The equations are solved in a non-spherical geometry that follows the centrifugal distortion of the star. Solutions are obtained using spectral elements and a discretisation of the star in `onion' shells, where the radial and the horizontal directions are respectively discretised using a Gauss-Lobatto and a Gauss-Legendre collocation grid. We thus have access to the latitudinal variations in all the fundamental parameters such as the effective temperature $T_{\rm eff}(\theta)$, the surface gravity $g(\theta)$, and the radius $R(\theta)$ thanks to the Legendre polynomial interpolation, where  $\theta$ is the colatitude.

\begin{figure}[t!]
\includegraphics[width=0.49\textwidth]{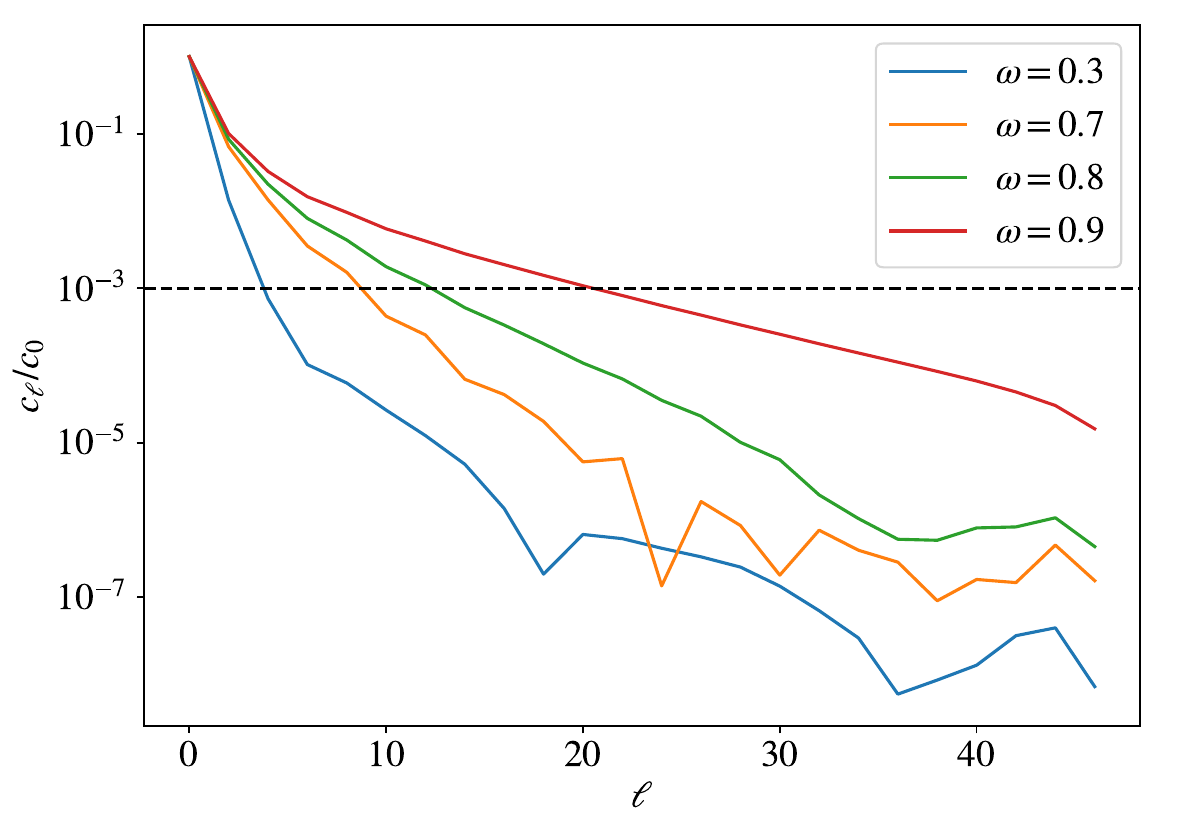}
\caption{Legendre spectra showing the rapid convergence of the expansion in spherical harmonics of the effective temperature given by the ESTER model. $\ell$ is the index of the spherical harmonic and $c_\ell$ the associated coefficient. We used a 3~$M_\odot$ model at various angular velocities. The dashed line shows the 0.1\% level of the truncation error.}
\label{C_n}
\end{figure}

Thanks to the spectral discretisation of the $\theta$ grid, accurate models can be computed with a small number of grid points. The spectral convergence of ESTER models in the angular dependence of $T_\mathrm{eff}$ is shown in Fig.~\ref{C_n}, where $\omega$ is the ratio of the actual equatorial rotation rate to the Keplerian rate at the equator. This figure shows that we need fewer than 22 spherical harmonics (or 11 grid points\footnote{The equatorial symmetry of the star implies that only spherical harmonics of even order are needed.}) to ensure a truncation error less than $10^{-3}$ at all rotation rates less than $\omega=0.9$. In Table~\ref{table:ntheta} we summarise the latitudinal resolution we used to minimise both spectral error and computing time.

\begin{table}
\begin{center}
\begin{tabular}{|c|c|}
\hline
$\omega$&$n_\theta$\\
\hline
$\leq$ 0.3 & 4\\
$0.3<\omega \leq 0.7$  & 8\\
$0.7<\omega \leq 0.9$ & 10\\
> 0.9 & $\geq$ 12\\
\hline
\end{tabular}
\end{center}
\caption{Number of latitudinal grid points $n_\theta$ that ensures a 0.1\% precision on $T_\mathrm{eff}$ and $\log g$ for various values of $\omega$.}
\label{table:ntheta}
\end{table}

\subsection{Spectrum synthesis}
\label{subsec:sp_synth}
Synthesising a spectrum requires a model atmosphere. Unlike the 1D case where effective temperature and gravity are constant over the stellar surface, we have to compute several atmosphere models following the local surface flux and gravity provided by the internal structure at each point of the visible surface of the star. Each atmosphere model is used to calculate a contribution to the observed spectrum. This  approach is  used in \cite{bouchaud+20} with the PHOENIX spectra from the G\"ottingen database \citep{Husser2013}. However, using pre-computed spectra requires   applying a weighting to account for the limb darkening effect \citep{Claret00}. In the present study we use the PHOENIX code \citep{hauschildt+99} to compute all the model atmospheres and corresponding spectra from scratch, which provides us with the intensity spectra for different directions, the limb darkening effect being included inevitably. To retrieve the specific intensity for any direction, we used a Legendre polynomial defined on a four-point Gauss-quadrature grid in $\mu=\cos \gamma$, where $\gamma$ is the angle between the normal to the surface and the direction of interest. We checked that such a small Gauss grid was able to ensure a numerical precision better than 5\% everywhere (actually better than 1\% in most cases). 

\begin{figure}[t!]
\includegraphics[width=0.49\textwidth]{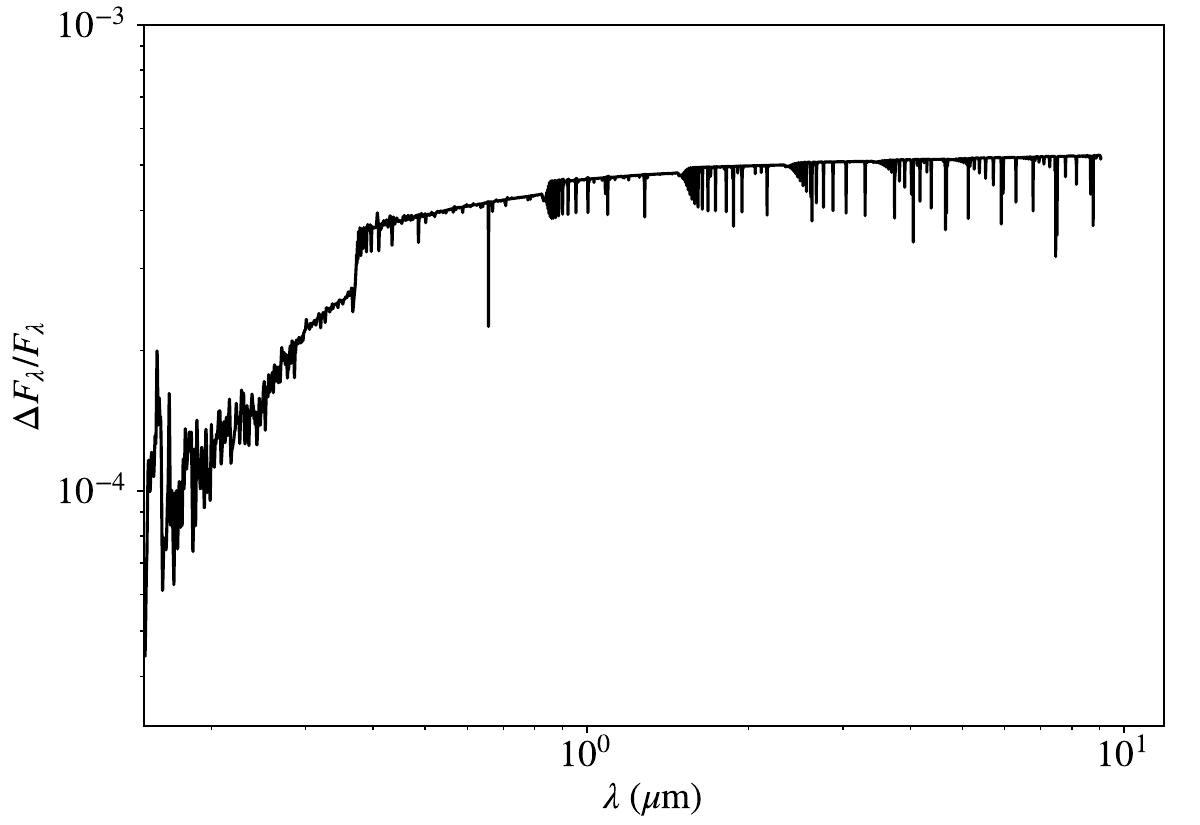}
\caption{Relative radiative flux difference between a plane-parallel and a spherical PHOENIX model with $T_\mathrm{eff}=20,000~\mathrm K$, $\log g=4$, and a radius of $2~R_{\odot}$, which is the smallest curvature radius for a $M=3~M_\odot$, $\omega=0.7$ stellar model.}
\label{sph_vs_pp}
\end{figure}

The model atmospheres and the spectra are calculated with the conservative microturbulence value of 2~km/s and the same set of abundances as the stellar structure, namely the chemical composition of \cite{AGSP09} with the meteoritic values of \cite{lodders+09} for the refractory elements, as suggested by \cite{Serenelli2010}.
Convection is treated with the mixing-length theory using a ratio of 1.25 for mixing length to pressure scale height. Since the relative radiative flux difference between a plane-parallel and a spherical model is below 0.1~\% in the wavelength interval we use (see Fig.~\ref{sph_vs_pp}), we performed  all the calculations with the plane-parallel geometry to reduce the computing time without impairing the results. Local thermodynamic equilibrium (LTE) is assumed for the model atmosphere and the spectrum synthesis.

\subsubsection{Geometric grid of the visible stellar surface}
The method we adopted to discretise the stellar surface is close to that described in \cite{bouchaud+20} and ensures that all the surface elements share roughly the same area. The star is first sliced along a meridian into $N_\theta$ latitudinal strips of constant angular width $\Delta \theta=\pi/2N_\theta$, where $N_\theta$ is  the number of atmosphere strips and $n_\theta$ is used to converge the internal structure, and are   independent. Keeping the size of each surface element roughly constant implies that the number of longitudinal elements varies from one latitudinal strip to the other. The typical size $\Delta S_{\!0}$ of a surface element is set in the strip surrounding the pole (of mean colatitude $\theta_0$) so as to have $N_{\phi_0}=N_{\phi}(\theta_0)$ surface elements. The parameter $N_{\phi_0}$ is chosen to have a spectrum as smooth as possible and we show below how its value is determined. To have the Doppler shifts symmetric compared to the rest frame, the surface elements are located symmetrically with respect to the meridian that contains the line of sight. Then, for any other strip of mean colatitude $\theta_k$, the number of longitudinal points $N_{\rm \phi}(\theta_k)$ is set to have $\Delta S(\theta_{k})$ as close as possible to $\Delta S_0$. The quantity $\Delta S(\theta_{k})$ is defined as
\begin{equation}
\label{ds_eq}
    \Delta S(\theta_{k})=r^2(\theta_k) \sqrt{1 + \frac{r_{\theta}^2}{r^2}} \sin{\theta_k}\Delta \theta \Delta \phi
,\end{equation}
where $r$ is the distance to the star  centre, $r_{\theta}=\partial r /\partial \theta$, and $\Delta \phi=2\pi/N_\phi(\theta_k)$ is the angular step in longitude. The visible surface is then obtained by selecting elements with $\mu(\theta_{k},\phi_{j})>0$, with $k\in[0,N_\theta -1]$ and $j\in[1,N_\phi(\theta_{k})]$. Thus,  $\mu(\theta_{k},\phi_{j})$ reads
\begin{eqnarray}
\label{mu}
     \mu(\theta_{k},\phi_{j})=\frac{1}{\sqrt{r^2 + r_{\theta}^2}}\left[(r\sin\theta_{k}-
     r_{\theta}\cos\theta_{k})\cos\phi_{j}\sin i \right. \nonumber \\ 
     +\left.(r\cos\theta_{k} + r_{\theta}\sin\theta_{k})\cos i\right]
,\end{eqnarray}
where $i$ is the angle between the line of sight and the rotation axis.

\subsubsection{Spectrum in the observer's frame}

At this stage, for each visible surface element located at $(\theta_k,\phi_j)$, the monochromatic contribution $\Delta I_\lambda(\theta_k,\phi_j)$ to the observed spectrum at wavelength $\lambda$ is defined as
\begin{equation}
    \Delta I_\lambda(\theta_k,\phi_j)=I_\lambda(T_\mathrm{eff}(\theta_k),g(\theta_k),\mu(\theta_k,\phi_j))\times \Delta S_\mathrm{proj}(\theta_k,\phi_j)
,\end{equation}
where $I_\lambda(T_\mathrm{eff}(\theta_k), g(\theta_k),\mu(\theta_k,\phi_j))$ is the specific intensity along the line of sight of a model atmosphere computed with $(T_\mathrm{eff}(\theta_k), g(\theta_k))$ and  $\Delta S_\mathrm{proj}(\theta_{k})$ is the surface element $\Delta S(\theta_{k})$ located at $\phi_j$ and projected on the sky. The observed spectrum is then the sum of each of these contributions, Doppler-shifted according to the motion along the line of sight of the surface element in question.

\begin{figure}[t!]
\includegraphics[width=0.49\textwidth]{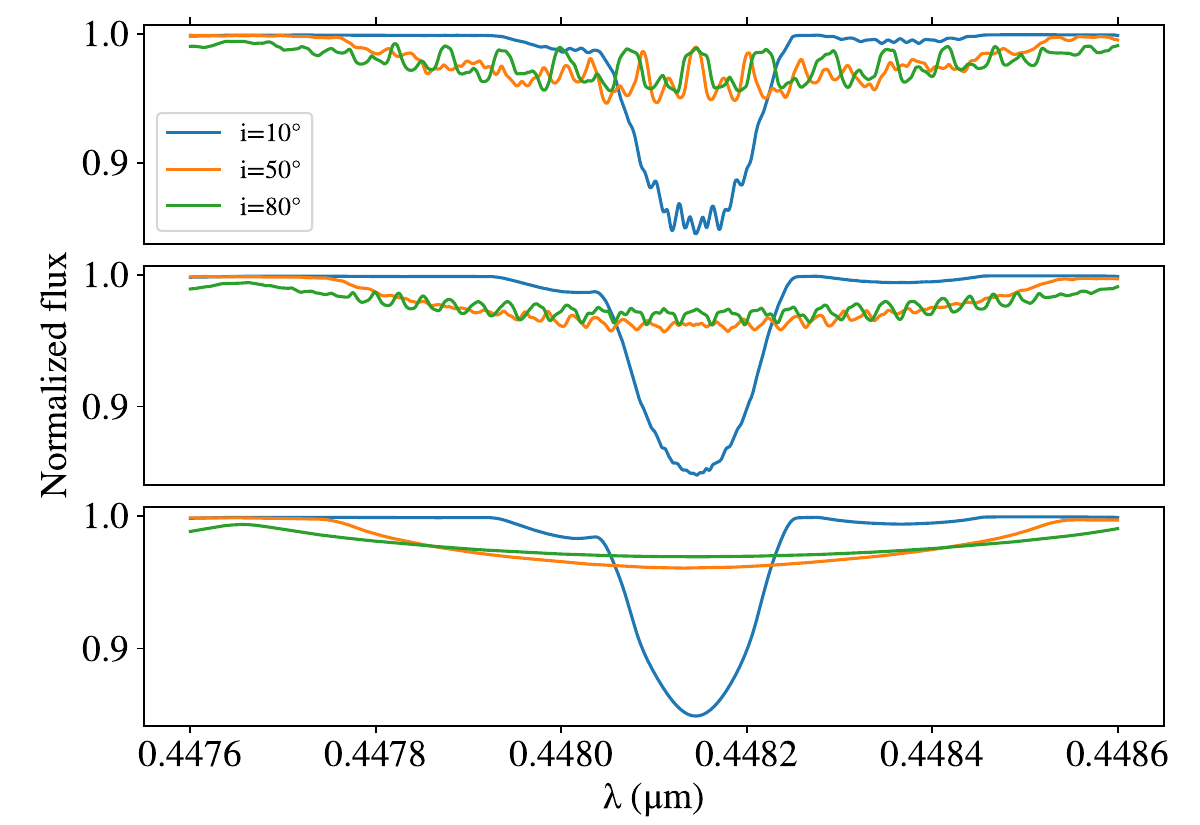}
\caption{Synthesis of the MgII $\lambda4481$ line of a $5~M_\odot$ star with $\omega=0.6$ at three angular resolutions: $(N_{\theta},N_{\phi_0})$=(10,10), (20,10), and (100,20) (top, middle, and bottom panels, respectively). In each panel the blue, orange, and green curves denote an inclination of $10^\circ$, $50^\circ$, and $80^\circ$, respectively.}
\label{graph_rbld_grid}
\end{figure}

Figure~\ref{graph_rbld_grid} shows a close-up of the resulting synthetic spectrum for a $5~M_\odot$ star with $\omega=0.6$ around the MgII $\lambda4481$ line, computed for three different inclinations of the rotation axis, various sets of $(N_\theta,N_{\phi_0})$, and a 1~pm wavelength step. The spectra are normalised to the continuum level, which was   computed in the same way, but without the line opacities. The wavy aspect of the spectra, especially visible at high inclinations, is due to the discretisation of the stellar surface. This numerical noise becomes negligible by setting $(N_{\theta},N_{\phi_0})$ to (100,20) or higher, as shown in Fig.~\ref{graph_rbld_grid}.

\section{Selected photometric quantities}
\label{section:influence}

The flux received by an observer depends on the effective temperature and effective gravity latitudinal distributions together with the inclination of the rotation axis on the line of sight. However, \cite{ELR11}  showed that the flux and the effective gravity are tightly related in the radiative envelope of early-type stars. Thus, the flux and the effective gravity distributions are both governed by the shape of the star. Hence, if we consider main sequence stars at a given evolutionary stage, we may expect that the energy flux towards an observer basically depends on two fundamental parameters of the star, its mass $M$ and its centrifugal flattening $\varepsilon$, and on the angle $i$ between its rotation axis and the line of sight. Instead of the centrifugal flattening $\varepsilon$, we  use $\omega$, the ratio of the equatorial angular velocity to the Keplerian angular velocity at equator. In the Roche model $\omega$ and $\varepsilon$ are simply related by $\varepsilon=\omega^2/(2+\omega^2)$, which is very close to the relation verified by the more realistic ESTER models. 

These three   quantities are expected to influence the shape of the star's spectrum, even more so when the star rotates rapidly. We therefore investigate the possibility that two colour indices together with a temperature expressing the surface heat flux may be able to constrain $i$, $\omega$, and $M$, if the age is given.

To this end, we chose to focus on the ZAMS, with solar metallicity. We
built a grid of models with masses from 2 to 7~$M_\odot$ and rotational rates $\omega$ between 0.3 and 0.8. The rotation axis inclinations ranged from 0 to $90^\circ$ with a $5^\circ$ step. The spectra were computed from 10~nm\ to $9,000$~nm with a 0.1~nm\ step.

\subsection{The infrared flux method}

Since we need an effective temperature representative of the surface heat flux of the star, we resort to that derived with the well-known IRFM, which was first devised by \cite{blackwell+77} to yield angular diameters along with effective temperatures. As a follow-up, \cite{blackwell+80} introduced the ratio $R=F/F_{\lambda_\mathrm{IR}}$ of the bolometric flux $F$ to the infrared monochromatic flux $F_{\lambda_\mathrm{IR}}$ to find effective temperatures. They computed theoretical values of this ratio $R_\mathrm{th}$ for a set of $T_\mathrm{eff}$ and several wavelengths $\lambda_\mathrm{IR}$. Then, for an observed value of $R$, the  IRFM temperature $T_\mathrm{IRFM}$ was retrieved by interpolation in the table linking $R_\mathrm{th}$ and $T_\mathrm{eff}$.

\begin{figure}[t!]
\includegraphics[width=0.49\textwidth]{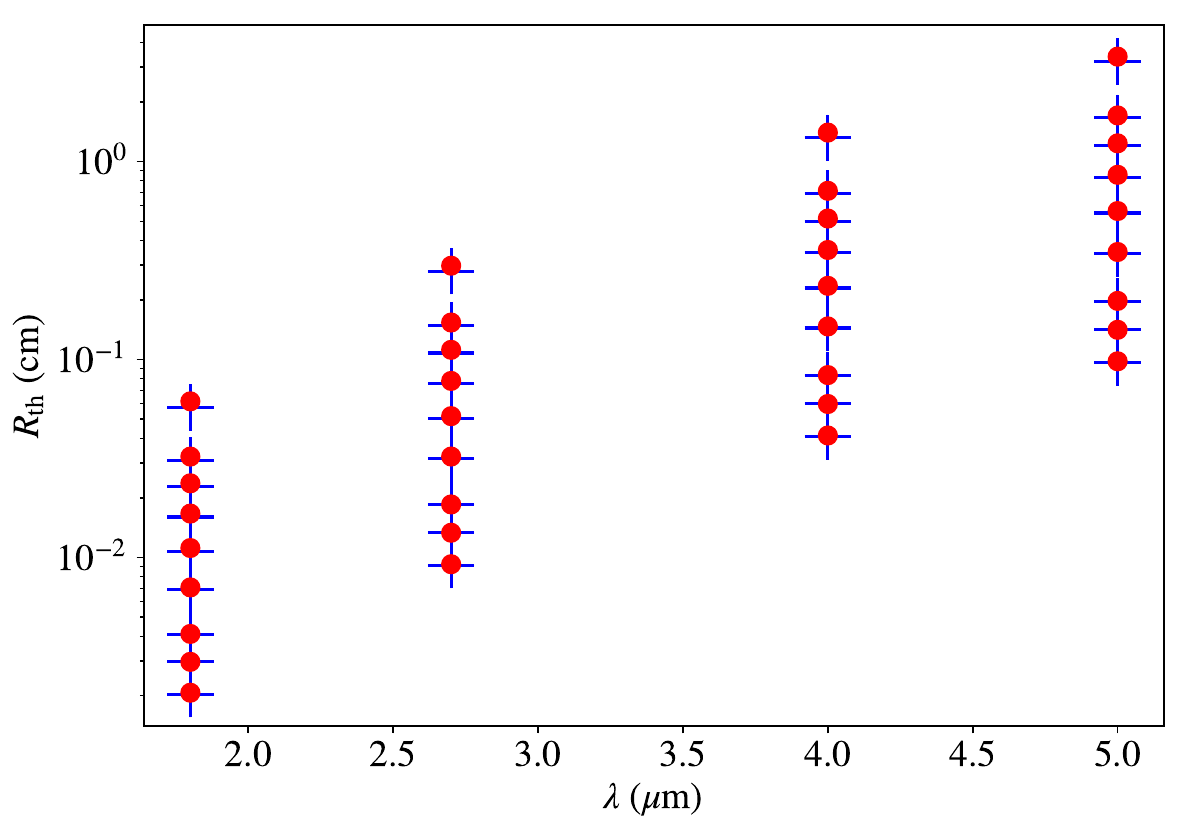}
\caption{Values of $R_\mathrm{th}$  at different wavelengths for $\log\ g=4$. The present data are plotted with blue crosses and those of \cite{blackwell+80} with red dots. For each wavelength, the effective temperature increases with increasing $R_\mathrm{th}$.}
\label{blackwell}
\end{figure}

We computed values of $R_\mathrm{th}$ with PHOENIX atmosphere models with $8,000~\rm K\leq T_\mathrm{eff}\leq 25,000$~K and $\log g=4.0$, which are typical values for the kind of stars we study.  The bolometric flux is obtained by an integration between the above-mentioned boundaries, the relative difference with larger intervals being less than $10^{-5}$. The results are shown in Fig.~\ref{blackwell} along with the original values of \cite{blackwell+80} for $\log g=4$. For similar temperatures and given $\lambda_\mathrm{IR}$, our values of $R_\mathrm{th}$ and those of Blackwell et al. are almost superimposed, showing that the relation between $T_\mathrm{eff}$ and $R_\mathrm{th}$ depends very weakly on the model atmosphere and the set of abundances, as already stated by \cite{blackwell+80}.

\begin{figure}[t!]
\includegraphics[width=0.49\textwidth]{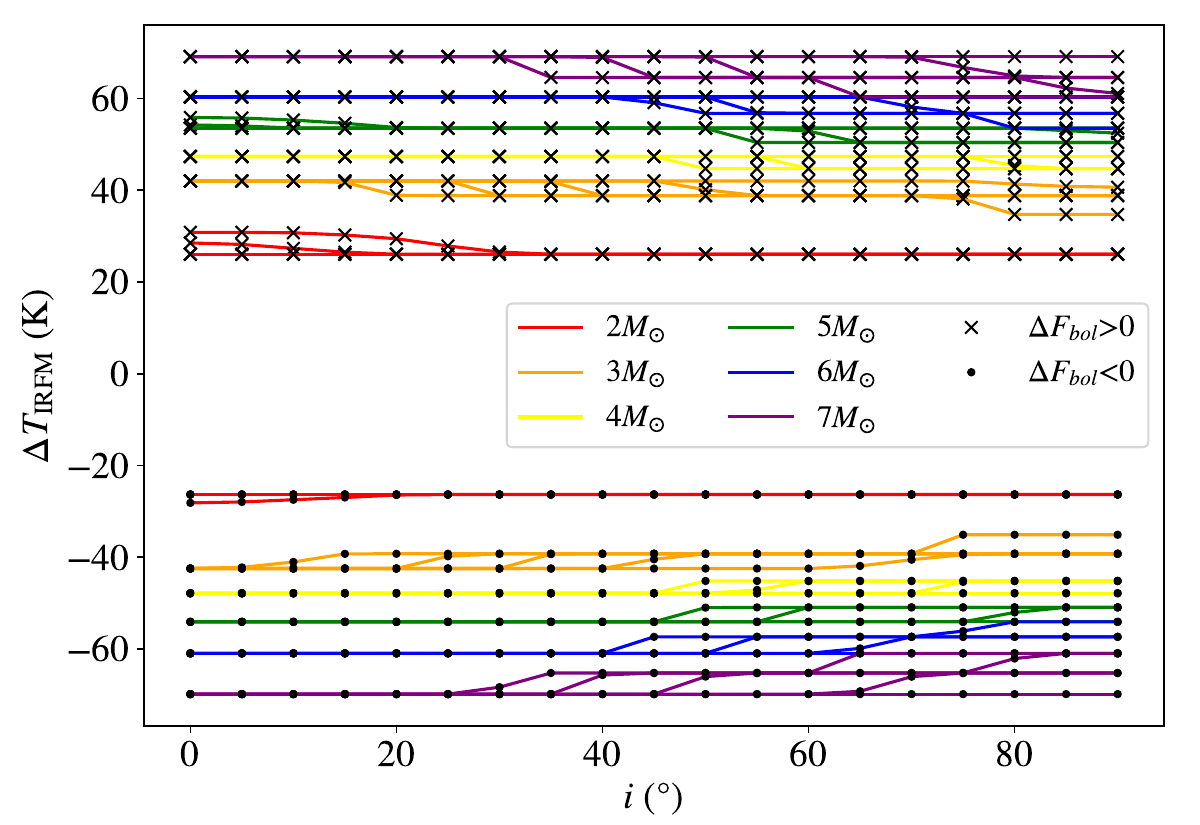}
\caption{Variation in $T_\mathrm{IRFM}$ induced by a 1\% variation in the bolometric flux for our grid of  ESTER/Phoenix models.
 For each mass, several rotation rates $\omega$ were considered, ranging from 0.3 to 0.7 with a   0.1 step.}
\label{delta_T}
\end{figure}

We then investigate the sensitivity of $T_\mathrm{IRFM}$ to measurement errors of the flux. For that we introduce a 1\% variation to the bolometric flux and monitor the resulting variations in $T_\mathrm{IRFM}$ on our grid of  ESTER/Phoenix models  with various rotation rates.  As shown in Fig.~\ref{delta_T}, the induced variation in $T_\mathrm{IRFM}$ in absolute value is 70~K at most. A similar 1\% variation introduced to the flux at a selected IR wavelength  has almost no effect on $T_\mathrm{IRFM}$. Hence, the error on $T_\mathrm{IRFM}$ essentially comes from that on the bolometric flux.

\begin{figure}[t]
\includegraphics[width=0.49\textwidth]{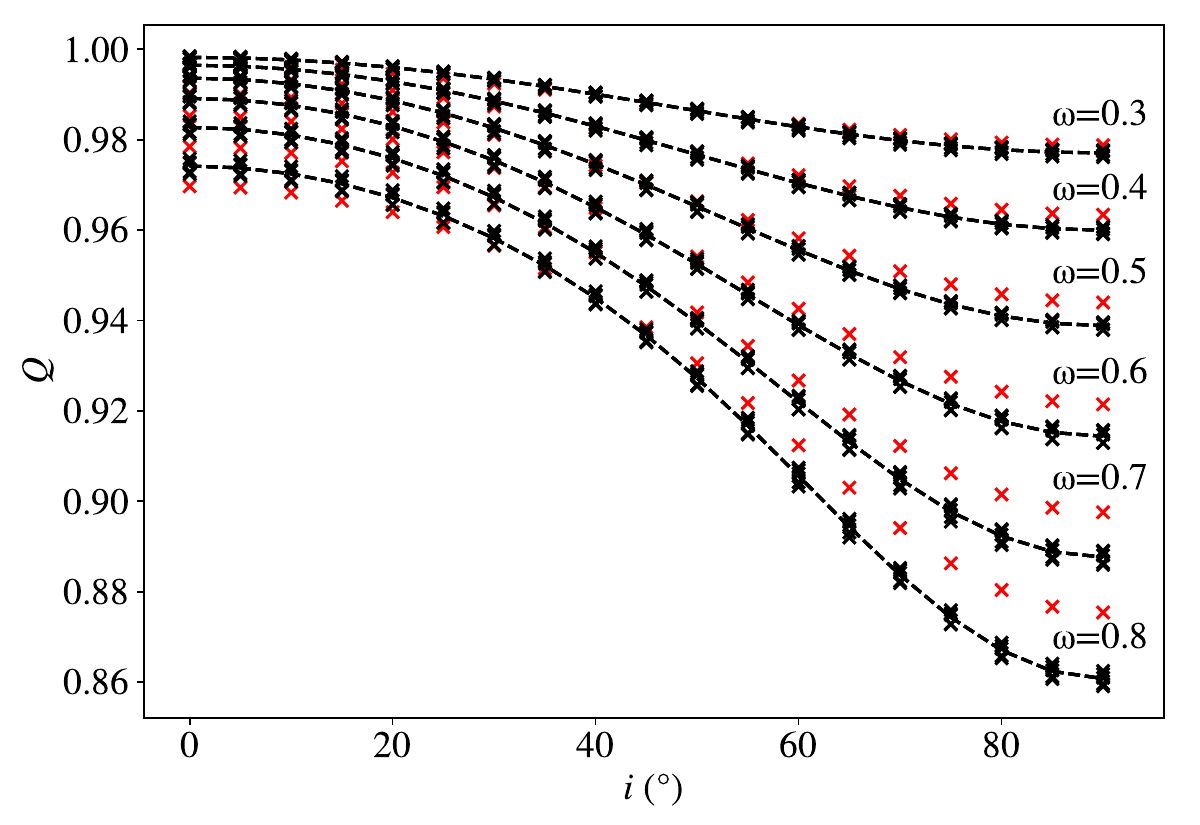}
\caption{Values of $Q$ for the models of our grid (crosses) vs inclination. Each dashed line corresponds to their average value over masses ($M\in$ [3-7] $M_{\odot}$) for a given rotation rate, ranging from $\omega$=0.3 (top) to $\omega=0.8$ (bottom). The red crosses show the 2~$M_\odot$ models.}
\label{Rap_IRFM_Xc1}
\end{figure}

We next investigate the sensitivity of $T_\mathrm{IRFM}$ to the geometry of the star, which is determined by the pair $(\omega,i)$. To this end, we consider the ratio
\begin{equation}
Q=\frac{T_\mathrm{IRFM}(M,\omega, i)}{T_\mathrm{IRFM}(M,\omega=0)}
\label{Rap_IRFM}
,\end{equation}
namely the IRFM temperature scaled by that  of the associated non-rotating model of the same mass. Quite nicely, the ratio $Q$ is almost independent of the mass if $M\geq3$~$M_\odot$, as shown by Fig.~\ref{Rap_IRFM_Xc1}. Hence, the IRFM temperature of a centrifugally distorted star viewed under an inclination angle $i$ is that of the corresponding non-rotating star corrected by a `shape factor' $Q\raisebox{-.7ex}{$\stackrel{<}{\,\sim\,}$}1$ actually describing the impact of gravity darkening on the flux received by the observer. The subsequent approximate relation
\begin{equation}
    T_\mathrm{IRFM}(M,\omega,i)\simeq T_\mathrm{IRFM}(M,\omega=0)\times Q(\omega,i)
\label{Eq_IRFM_Q}
,\end{equation}
{where $Q(\omega,i)$ are bi-cubic splines fitting the $Q$ values,} can be very useful to monitor the impact of rotation and inclination on main sequence evolutionary tracks computed with non-rotating 1D models.

This   independence of the Q-factor with respect to mass seems to disappear below 3~$M_\odot$\ (e.g. $2~M_\odot$ models, red crosses in Fig.~\ref{Rap_IRFM_Xc1}). The dependence is not strong but notable, especially at high angular velocity and high inclination, for an as-yet-unknown reason.

\subsection{The Str\"omgren index $c_1$ }\label{present_c1}

\begin{figure}[t!]
\includegraphics[width=\linewidth]{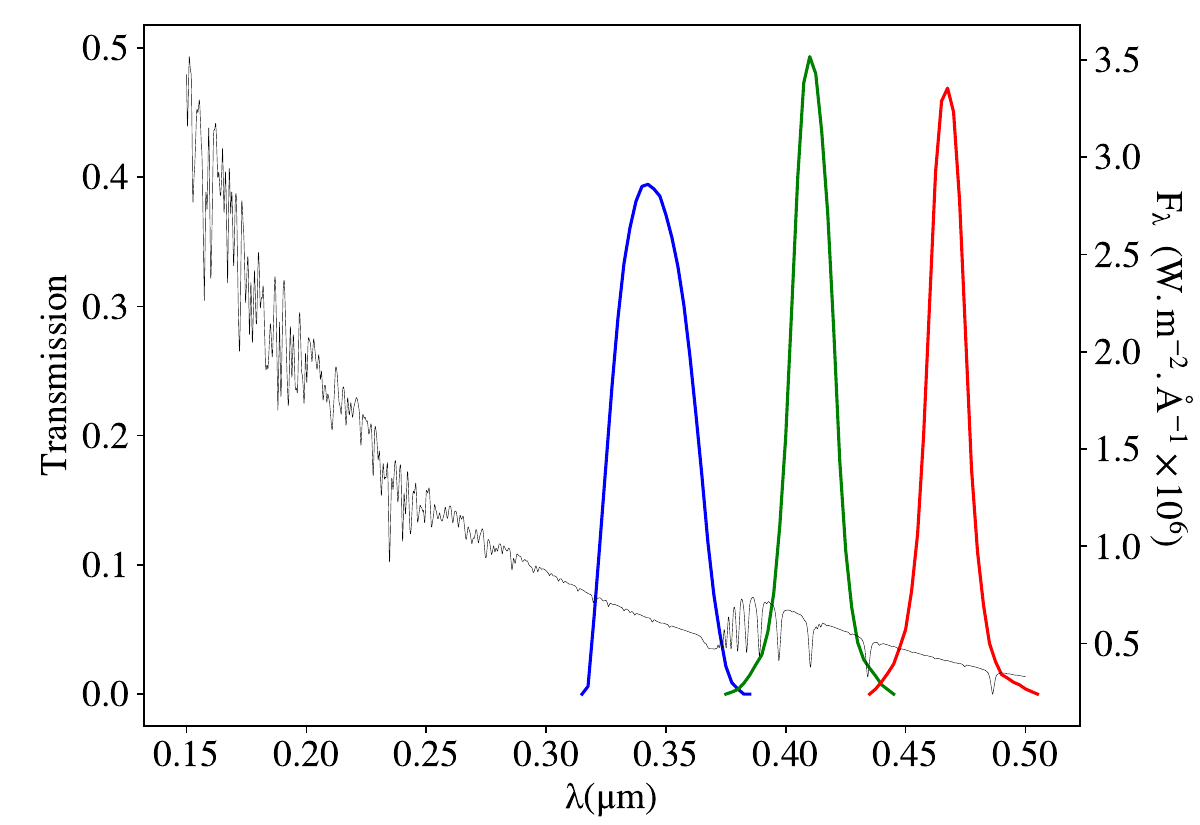}
\caption{Transmission curves of the $u$, $v$, and $b$ filters of the  Str\"omgren photometric system (blue, green, and red curves, respectively). The underlying spectrum (black line) comes from a model with  $M=5~M_{\odot}$, $\omega$=0.5, and $i=65^\circ$ located at 5 pc from Earth.}
\label{filters_index}
\end{figure}

The next spectral quantity that we consider is the Str\"omgren index $c_1$. \cite{Stromgren1963} defined a photometric system consisting of four intermediate-width bands $(u, v, b, y)$. The $c_1$ index (in magnitudes) is the  colour difference $(u-v)-(v-b)$. The transmissions of the filters entering the definition of $c_1$ are shown in Fig.~\ref{filters_index}. 

We use the \cite{Kurucz1993} program to compute synthetic $c_1$ 
indices from our spectra. To allow direct comparisons with observed photometry, this program takes into account the transmission of the filters, the transmission of the Earth's atmosphere, and the response of the detector. The zero point is set by equating the $c_1$ value, which we obtained with the spectrum of our Vega model (see Table~\ref{Vega_param}), to the observed value of \cite{Crawford_Barnes1970}.

\begin{figure}[ht!]
\centering\includegraphics[width=\linewidth]{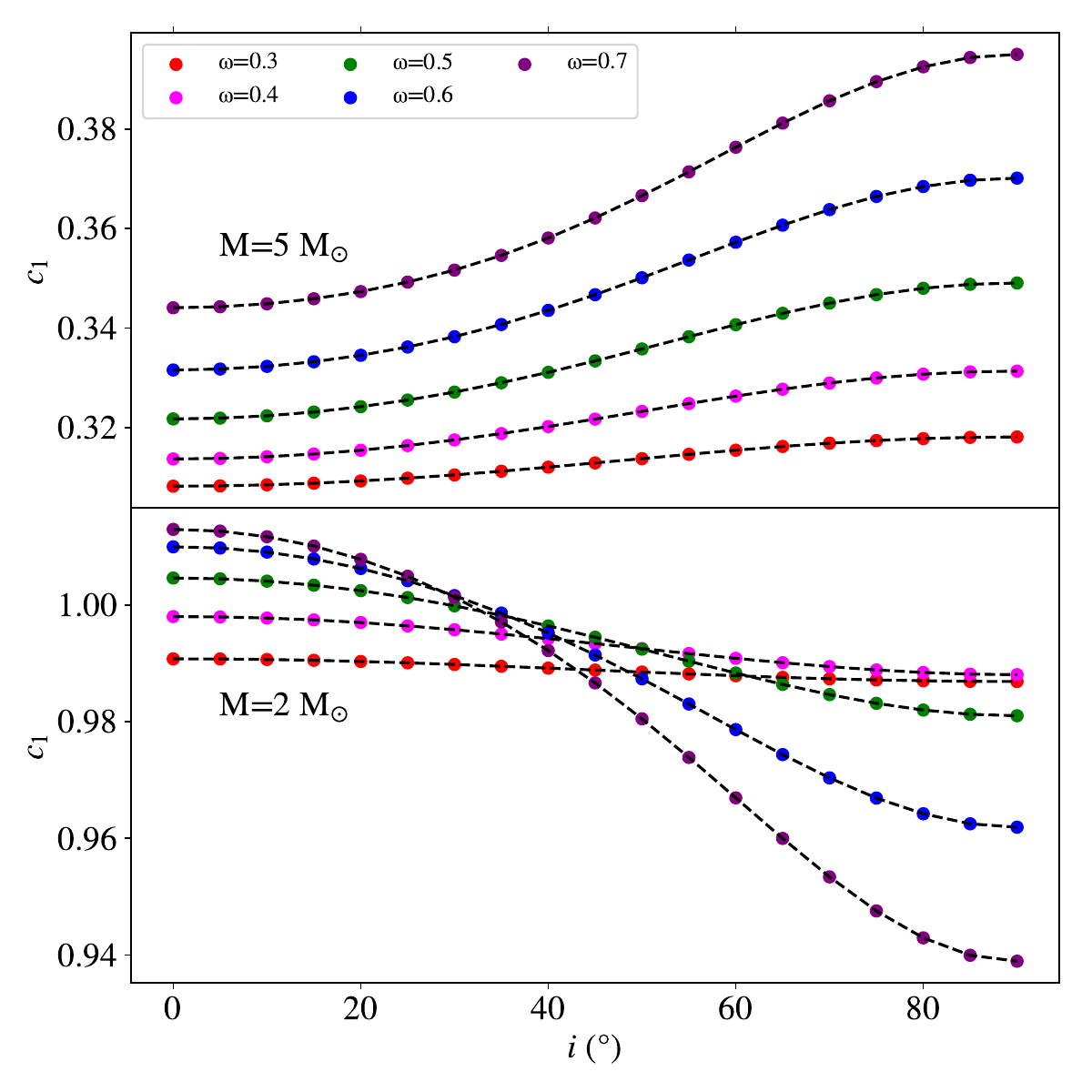}
\caption{Variations in the $c_{1}$ index with the inclination $i$ for several values of $\omega$ in the range [0.3;0.7] with a  step of 0.1, for a $5~M_{\odot}$ model (upper panel) and for a $2~M_\odot$ model (lower panel). The dashed curves represent the corresponding fits.}
\label{c1_2-5}
\end{figure}

The  variations in $c_1$ with $T_\mathrm{eff}$ and $\log g$ of spherical models are illustrated in Figs.~2 and 3 of \cite{Moon_Dworetsky1985}. For stars hotter than about 9,500~K, at given $\log g$, $c_1$ is an indicator of $T_\mathrm{eff}$, which increases with decreasing $c_1$. A small dependence versus $\log g$ also exists. In the case of a fast rotating star, $\log g$ and $T_\mathrm{eff}$ decrease from pole to equator and both variations increase $c_1$. The overall effect with respect to the rotation axis inclination for such stars is plotted in the top panel of Fig.~\ref{c1_2-5} for a 5~$M_\odot$ model: the lowest value of $c_1$ is met when the star is seen pole on.

For temperatures less than 8,500~K, this effect reverses: $c_1$ decreases from pole to equator, as illustrated in the bottom panel of Fig.~\ref{c1_2-5} with a 2~$M_\odot$ model. In this  case the temperature effect overcomes the $\log g$ effect. As can be guessed, between these two temperatures, the dependence of $c_1$ with inclination is weaker and can even be non-monotonic.

\subsection{Hubble Space Telescope colour index $c_2$}

\begin{figure}[t!]
\includegraphics[width=\linewidth]{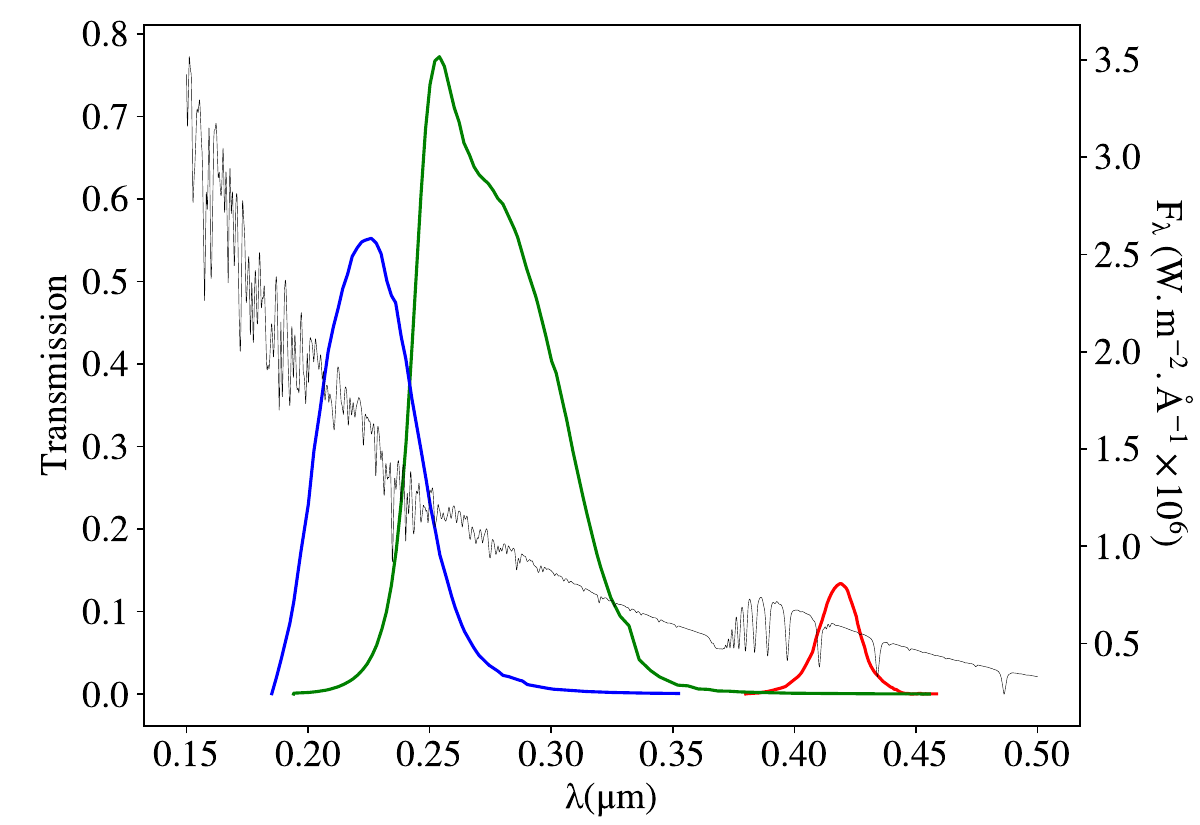}
\caption{Same as Fig.~\ref{filters_index} for the three HST filters F220, F250, and F419 (blue, green, and red curves respectively) used to compute the $c_2$ index.}
\label{filters_index_HST}
\end{figure}

To probe the shape of our spectra at wavelengths shorter than the Balmer jump, we built a new index $c_2$ in a similar way to $c_1$. The $c_1$ index can be interpreted as a measure of the curvature of the spectrum at the low-frequency end of the Balmer jump. Thus, we built $c_2$ to measure the curvature of the spectrum at the high-frequency end of the Balmer jump. Thus, we need two filters on the blue side of the Balmer jump. For this new index we chose three filters from the Hubble Space Telescope filter set, namely HSP\_VIS\_F419N\_B, ACS\_HRC\_F250W, and HSP\_UV1\_F220W\_A,
respectively referred to as F419, F250, and F220 in the following. Other choices are  possible as long as the curvature of the spectrum is monitored. Thus, we define $c_2$ as 

\begin{equation*}
\label{expression_c2}
c_2=-\frac{5}{2}\left( \log \left( \frac{\int F_\mathrm{F220}}{\int F_\mathrm{F250}} \right)-\log \left( \frac{\int F_\mathrm{F250}}{\int F_\mathrm{F419}} \right)\right)\quad\mathrm{mag.}
\end{equation*}
The transmissions of these three filters are shown in Fig.~\ref{filters_index_HST}.

\begin{figure}[t!]
\includegraphics[width=\linewidth]{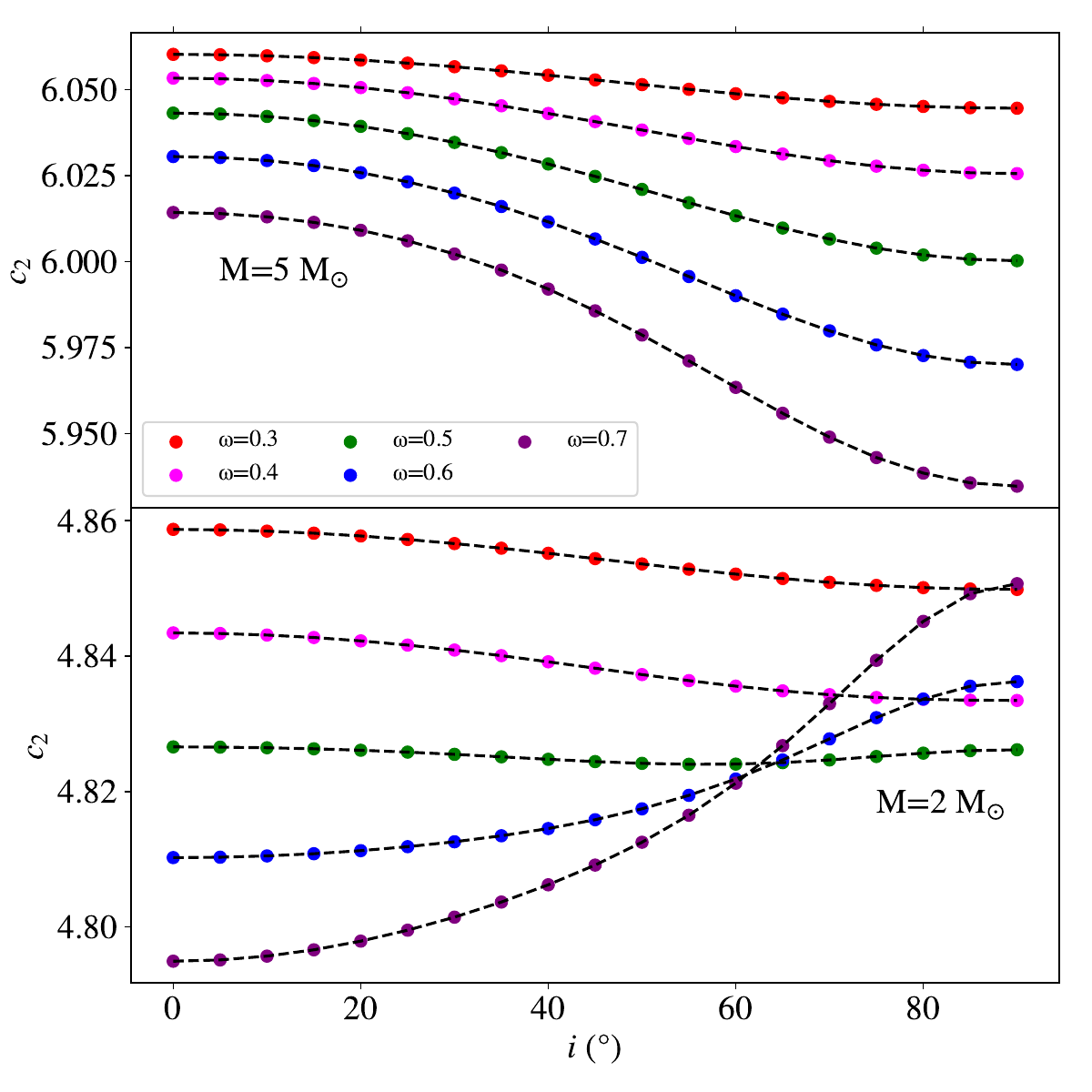}
\caption{Same as Fig.~\ref{c1_2-5}, but for the $c_{2}$ index.}
\label{c2_2-5}
\end{figure}

In Fig. \ref{c2_2-5} we show the variations in $c_2$ with inclination and rotation rate for a $2~M_{\odot}$ and a $5~M_{\odot}$ model. These variations are actually    opposite to  those of $c_1$, which is a good property for inversion matters.

\section{$\it M$, $\omega$, and $i$ determination from photometric observables}
\label{section:determination}

In the previous section we computed the three functions
\begin{equation}
    T_\mathrm{IRFM}(\it M,\omega,i),\quad c_1(\it M,\omega,i), \quad c_2(\it M,\omega,i)
    \label{les_trois_f}
\end{equation}
on a grid of triplets $(M,\omega,i)$ for ZAMS models.

The relation between the observed $T_\mathrm{IRFM}$, $c_1$, and $c_2$ and the parameters $(M,\omega,i)$ is non-linear, as shown by the various examples in Figs.~\ref{c1_2-5} and \ref{c2_2-5}. 
The exact correspondence between ($T_\mathrm{IRFM}$, $c_1$,  $c_2$) and $(M,\omega,i)$ is only known at the gridpoints. Since the computation of the three observables is indeed computationally very demanding, if we need to evaluate these three functions at any mass, rotation or inclination, we are forced to resort to interpolations by using
tri-cubic splines. The drawback is that these interpolations introduce some numerical noise, which perturbs somewhat the process to retrieve $(M,\omega,i)$ from ($T_\mathrm{IRFM}$, $c_1$,  $c_2$) as we show below.

\subsection{Initialisation}

\begin{figure}[!t]
\includegraphics[width=0.49\textwidth]{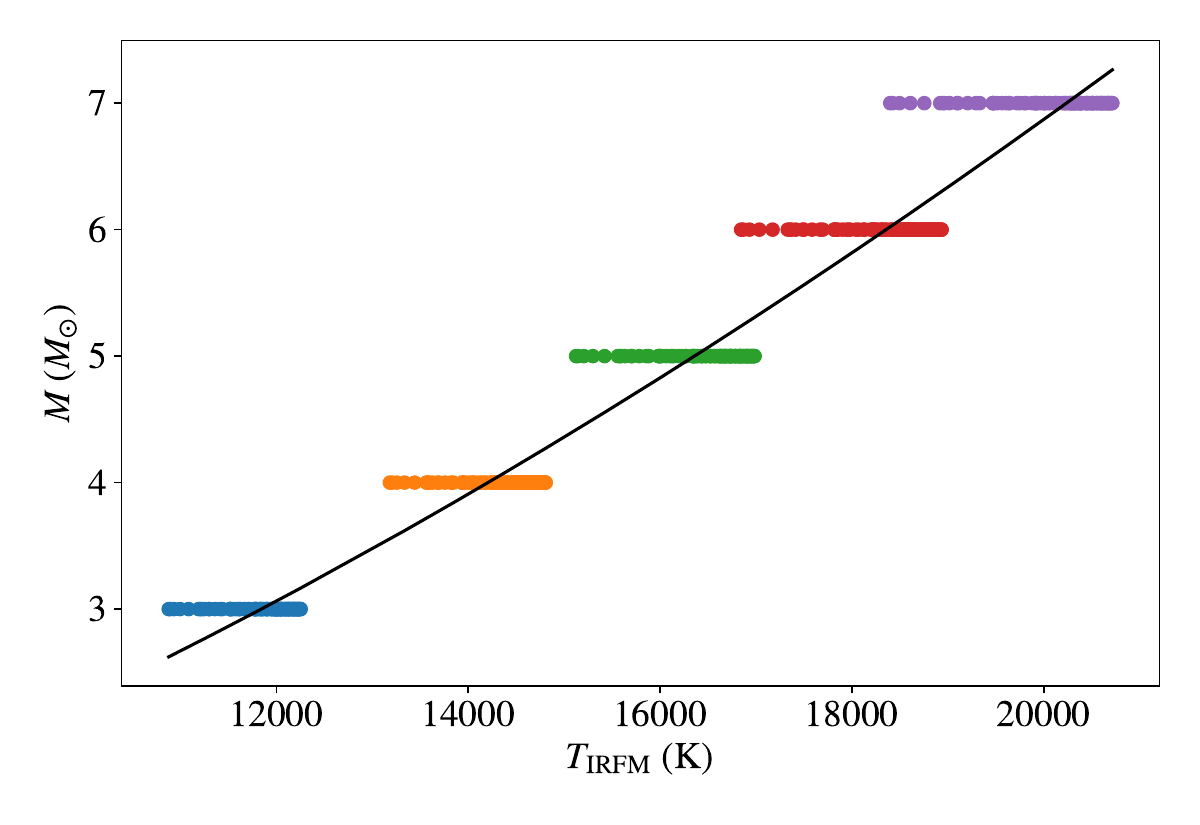}
\caption{Models of our grid in the $T_\mathrm{IRFM}$--mass plane where each dot represents a model of given $(\omega,i)$. The black solid curve shows the fit of the average temperature for each mass.}
\label{fit_MTirR}
\end{figure}

As for any resolution of a non-linear problem, we need  an initial guess not too far from the target solution to start the solver. Hence, we first need a reasonable guess of $\it M$.

Since our models have a fixed metallicity and age, their effective temperature, and thus $T_\mathrm{IRFM}$, is essentially controlled by their mass, as illustrated by Fig.~\ref{fit_MTirR}. Clearly $(\omega,i)$ impacts $T_\mathrm{IRFM}$, and the relation between $T_\mathrm{IRFM}$ and $M$ can only be approximate. We derived the following fit
\begin{equation}
M=0.96\left ( \frac{T_\mathrm{IRFM}}{5777} \right)^{1.58}M_{\odot}
\label{empir}
,\end{equation}
which is shown in Fig.~\ref{fit_MTirR} as a black line. This relation provides an approximate mass with a maximum error of 15\% for the worst case (i.e. when $M=2~M_{\odot}$, $\omega=0.3$, $i=90^\circ$). Hence, this relationship determines a set of possible values of $\it M$, while $\omega$ and $\sin i$ are still unconstrained.

\begin{figure}[t!]
    \includegraphics[width=\linewidth]{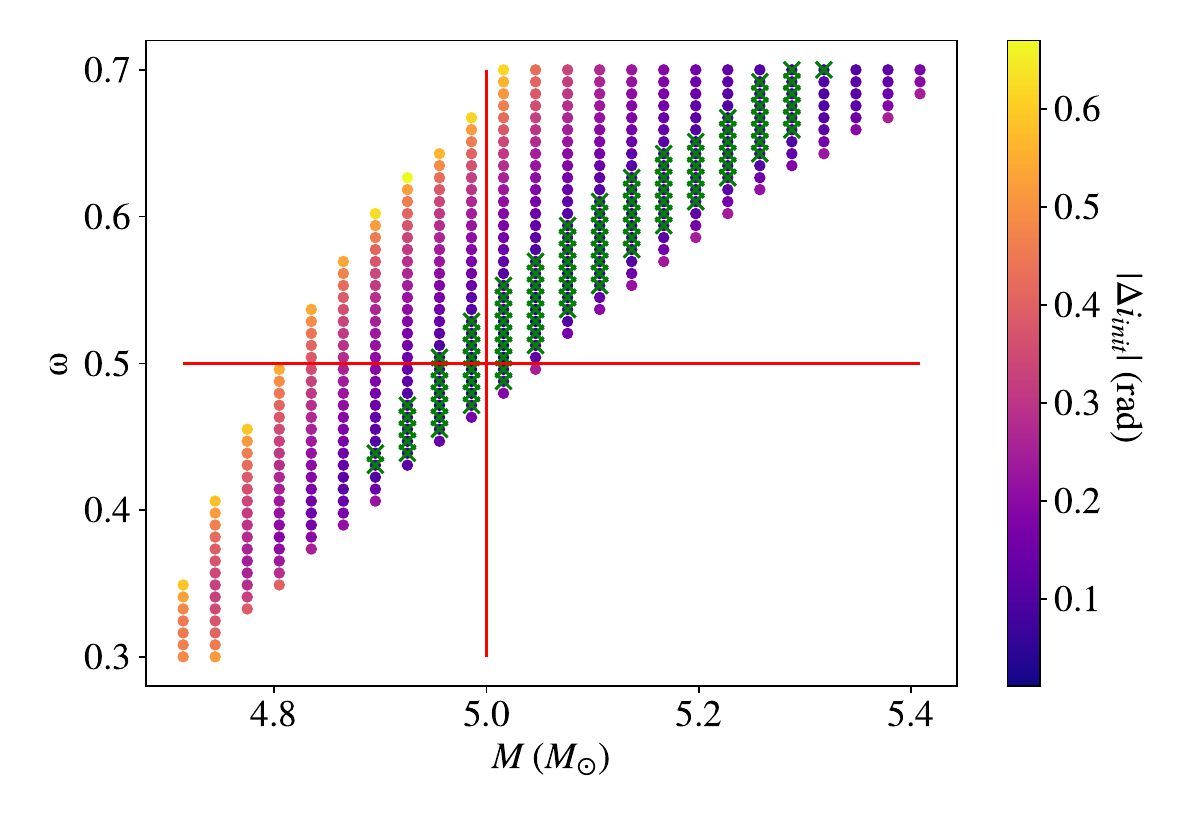}
    \caption{Example of a filtered grid (see text) for a 5~$M_{\odot}$ model with $\omega=0.5$ and $i=65^\circ$. The dots represent models that meet conditions~(\ref{filter}), colour-coded by the scattering between the inclinations obtained from each observable, as defined by the left-hand side of Eq.~\ref{Filter_conditions2}. Green crosses stand for initialisation models for which Eq.~\ref{Filter_conditions2} is verified with $\varepsilon=0.1$. The big red cross shows the location of our observed star.}
    \label{Filtrage 2}
\end{figure}

To limit the range of potential solutions, we first consider two series of models: pole-on models ($i=0^\circ$) and equator-on models ($i=90^\circ$). For these two inclinations, we determine the range of $\omega$ for which the three observed values of $(T_\mathrm{IRFM},c_1,c_2)$ are between the extreme values imposed at $i=0^\circ$ and $i=90^\circ$:
\begin{eqnarray}
\label{Filter_conditions1}
T_{\rm IRFM}(M,\omega,i =90^{\circ})\leq T_{\rm IRFM}^{\rm obs}\leq T_{\rm IRFM}(M,\omega, i=0^{\circ}), \nonumber \\
(c_1^{\rm obs}-c_1(M,\omega, i =90^{\circ}))\times(c_1^{\rm obs}-c_1(M,\omega, i =0^{\circ}))\leq 0, \label{filter} \\
(c_2^{\rm obs}-c_2(M,\omega, i =90^{\circ}))\times(c_2^{\rm obs}-c_2(M,\omega, i =0^{\circ}))\leq 0. \nonumber
\end{eqnarray}
Here the superscript `obs'  stands for the observed (i.e. target) values. The
result of this selection is illustrated in Fig.\ref{Filtrage 2}, where we show the values of $(M,\omega)$ that meet the above requirements, for an observed ZAMS model of $5~M_{\odot}$ at $\omega=0.5$ and $i=65^\circ$.

Models complying with system (\ref{Filter_conditions1}) are still numerous, as shown in Fig.~\ref{Filtrage 2}. To further restrict the possibilities, we consider each possible solution ($M_k,\omega_k$) of this set (i.e. each dot in Fig.~\ref{Filtrage 2}) and solve for the inclination for each observable. In other words, we solve independently the three  equations with the Newton-Raphson algorithm,

\begin{subequations}
\begin{equation}
T_\mathrm{IRFM}^{\rm obs}-f_{T_\mathrm{IRFM}}(M =M_k,\omega=\omega_k,i)=0
\label{itirfm}
,\end{equation}
\begin{equation}
c_{1}^{\rm obs}-f_{c_1}(M =M_k,\omega=\omega_k,i)=0
,\end{equation}
\begin{equation}
c_{2}^{\rm obs}-f_{c_2}(M =M_k,\omega=\omega_k,i)=0,
\end{equation}
\end{subequations}
where $f_{T_\mathrm{IRFM}},f_{c_1}$, and $f_{c_2}$ are the tri-cubic splines. We thus obtain three values for the star's inclination for each possible pair $(M_k,\omega_k)$. Obviously, a suitable pair is one for which the three inclinations are equal. Due to the analytic fits, we cannot expect a strict equality and we select the acceptable pairs $(M_k,\omega_k)$ by imposing that there is an inclination $i_0$ such that

\begin{equation}
    \sqrt{(i(T_\mathrm{IRFM})-i_0)^2+ (i(c_1)-i_0)^2 +(i(c_2)-i_0)^2} \leq \varepsilon
    \label{Filter_conditions2}
,\end{equation}
where $i(T_\mathrm{IRFM})$ is the inclination which meets the $T_\mathrm{IRFM}$ constraint \eqref{itirfm} for $( M_k,\omega_k)$, and {mutatis mutandis} for $i(c_1)$ and $i(c_2)$. If such an inclination $i_0$ exists for a small enough value of $\varepsilon$, then
a model matching the observation exists as well. Numerical experiments showed that $\varepsilon$ should not be larger than $10^{-1}$ for the algorithm to converge to the target solution. This initial guess $(M_k, \omega_k, i_0)$ is then used to derive the final solution with a Levenberg-Marquardt algorithm. We show in Fig.~\ref{Filtrage 2} an example of the set of models that match Eq.~\ref{Filter_conditions2} (green crosses).

Some difficulties arise, however, and come from the fact that the three functions of Eq.~\ref{les_trois_f} are only known  thanks to interpolating splines, which reproduce the values of our grid of models with limited accuracy out of the grid points. Hence, some differences can appear between the  `true' values of $T_\mathrm{IRFM}$, $c_1$, and $c_2$ and those computed using the fits for model that are not on the grid points, so that we may not be able to retrieve the exact solution ($\it M$, $\omega$, $i$) in these cases.

\subsection{Robustness of our resolution}
\label{subsection:robustness}

\begin{figure}[t!]
    \includegraphics[width=\linewidth]{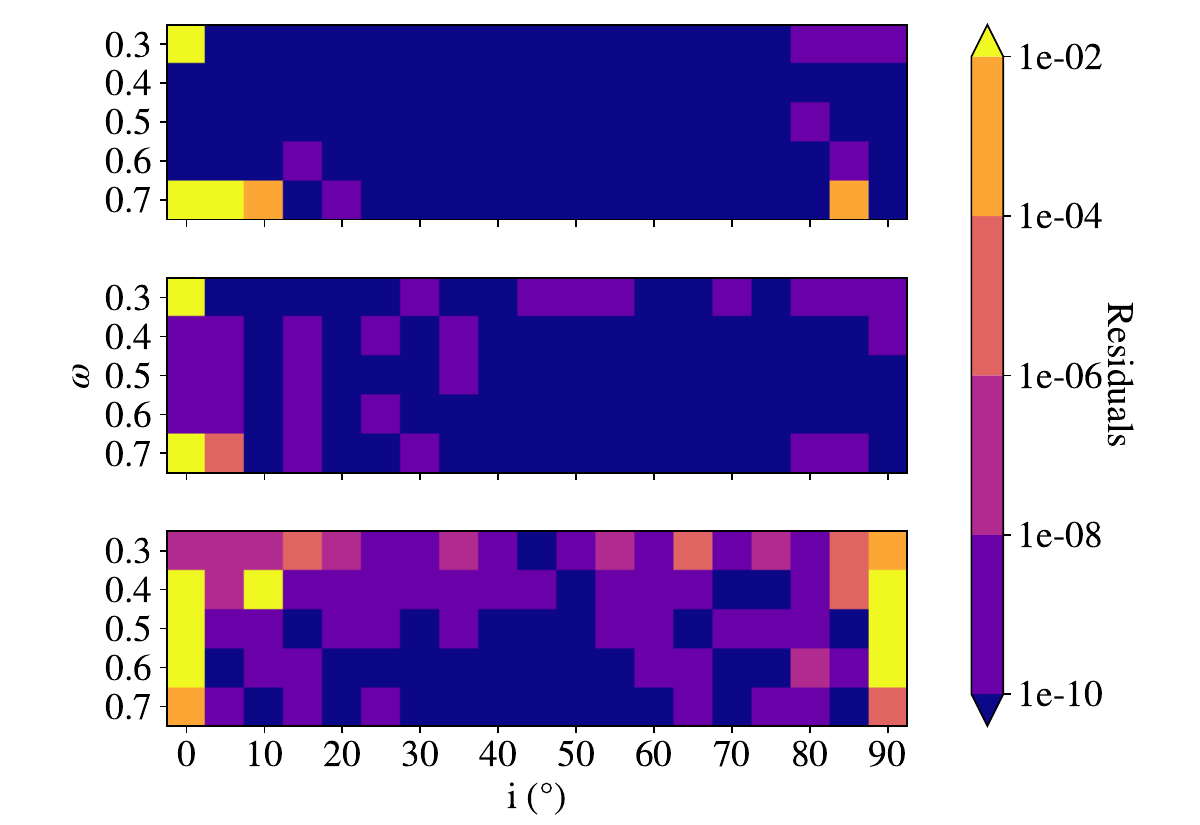} 
    \caption{Error maps in the $\omega$, $i$ plane, for stars with masses equal to 3, 5, and 7~$M_{\odot}$ (top, middle, and bottom panel, respectively). The colour stands for the absolute error between the retrieved and the original models defined as $\sqrt{(M_\mathrm{sol}-M_\mathrm{obs})^2+ (\omega_\mathrm{sol}-\omega_\mathrm{obs})^2 +(i_\mathrm{sol}-i_\mathrm{obs})^2}$, where $M$ is in $M_{\odot}$ and $i$ is in radians. The subscripts `sol' and `obs'  stand for the solution and the original model, respectively. The yellow cells correspond to differences greater than $10^{-2}$, where the resolution possibly failed.}
    \label{error_maps_clamped}
\end{figure}

Since the interpolating splines cannot give us an exact inversion, it is worth evaluating the error generated by the inversion 
\[
T_\mathrm{IRFM}, c_1, c_2 \longrightarrow M, \omega, i \; .
\]
To this end, we apply the inversion to all the observed quantities sampled on our gridpoints. In most cases the algorithm succeeds in retrieving the parameters of the model, but it fails sometimes, especially in the neighbourhood of the grid boundaries. In Fig.~\ref{error_maps_clamped} we show maps of errors in the $(\omega,i)$ plane for three masses of the grid. In most cases the parameters of the models are obtained with a negligible error.

\begin{figure}[t!]
    \centering
    \includegraphics[width=\linewidth]{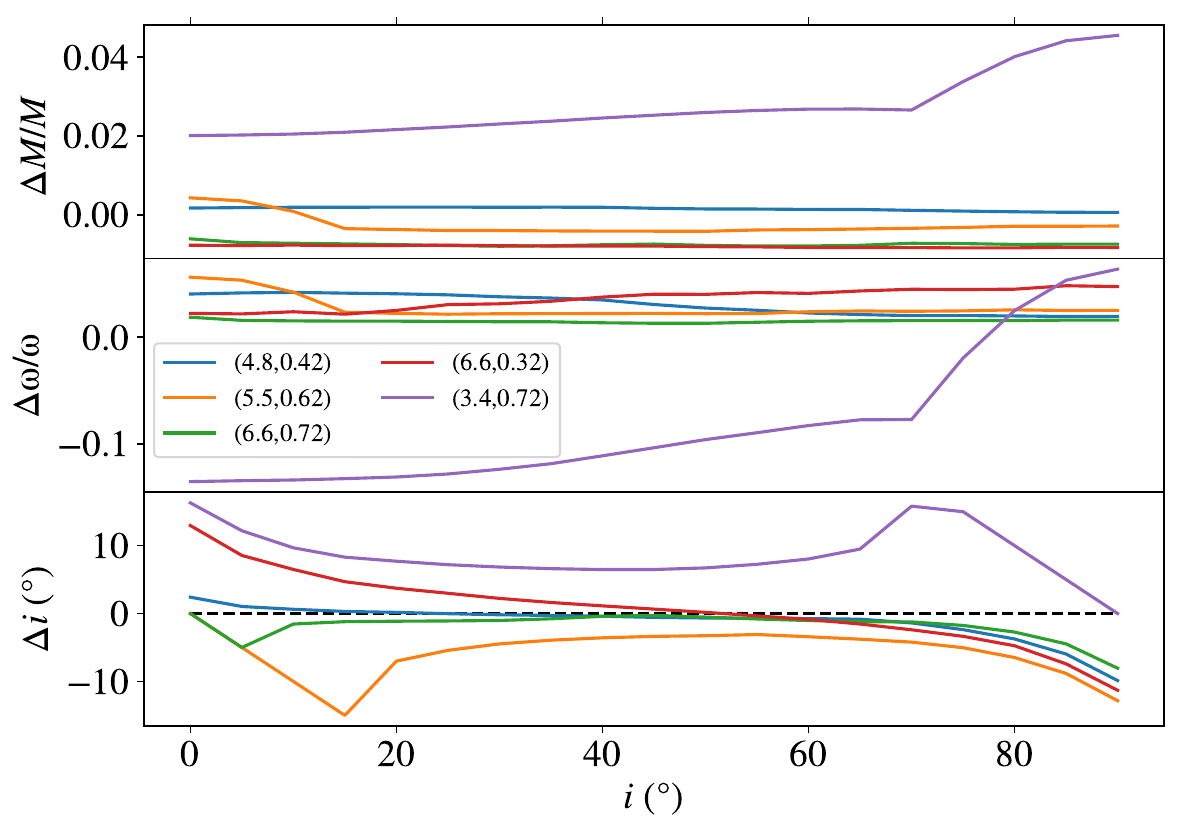} \caption{Errors on the determination of M (top), $\omega$ (middle), and $i$ (bottom) for models with $(M, \omega )$=(4.8, 0.42), (5.5, 0.62), (6.6, 0.72), (6.6, 0.32), and (3.4, 0.72) (blue, orange, green, red, and purple lines resp.), with $M$ in $M_\odot$. }
    \label{intergrid}
\end{figure}

The next test is to invert observables associated with models that are not on the gridpoints. In Fig.~\ref{intergrid} we show the error measured on five models that are representative of the general trend. Typically, the error strongly  depends on the mass range: it is  a few percent when $M\geq3.5~M_\odot$, but may exceed 10\% for lower masses. This sensitivity to interpolation errors is amplified by the weak variations in indices $c_1$ and $c_2$ for these masses. Inclination errors may reach $\pm15^\circ$ in the worst cases, namely for rotators viewed pole-on. Even though these errors may seem important, they can easily be reduced with the use of a finer grid, and focusing on a narrower range of parameters in the search process.

\section{Application to a real case: Vega}
\label{section:vega}
\subsection{Preliminaries}

Vega ($\alpha$ Lyrae) is a traditional standard of photometry \citep{bohlin+19}, and as such has been observed in many ways, especially with interferometers in recent years \citep{Aufdenbergetal06,petersonetal06a,monnier_etal12}. Using observations with the CHARA-array and MIRC beam combiner operating at $1.65~\mu$m, \cite{monnier_etal12} derived a concordance model that matches interferometric and spectroscopic constraints. Since the aim of our work is to derive fundamental parameters from spectroscopic indices, we naturally test our method on Vega.

As a first step, we derived a 2D-ESTER model that matches the most robust observational constraints. From the concordance model of \cite{monnier_etal12}, we assume an inclination of Vega rotation axis of $i=6.2^\circ$, which implies an equatorial velocity of 196~km/s using the spectroscopically derived value of $V\sin i=21.2$~km/s \citep{takeda_etal08}. We also adopt the polar and equatorial radii given by \cite{monnier_etal12} and the derived polar and equatorial effective temperatures. We match these observational data with a 2D-ESTER model of mass $M=2.374~M_\odot$ with metallicity $Z=0.0093$ \citep{yoon_etal08}, a hydrogen mass fraction of $X_{\rm env}=0.7546$ in the envelope, and $X_{\rm core}=0.2045$ in the convective core, as actually found by \cite{ELR13}. Hereafter, we   use instead $x_\mathrm c=X_{\rm core}/X_{\rm env}=0.271$ to denote the evolutionary stage of the model. All the parameters of our model are summarised in Table~\ref{Vega_param}.  We note that our mass is higher than that of the concordance model of \cite{monnier_etal12}. It may be a consequence of the Roche approximation used by these authors. This difference is however not critical for our test.

\begin{figure*}[t!]
    \centering
    \includegraphics[width=\linewidth]{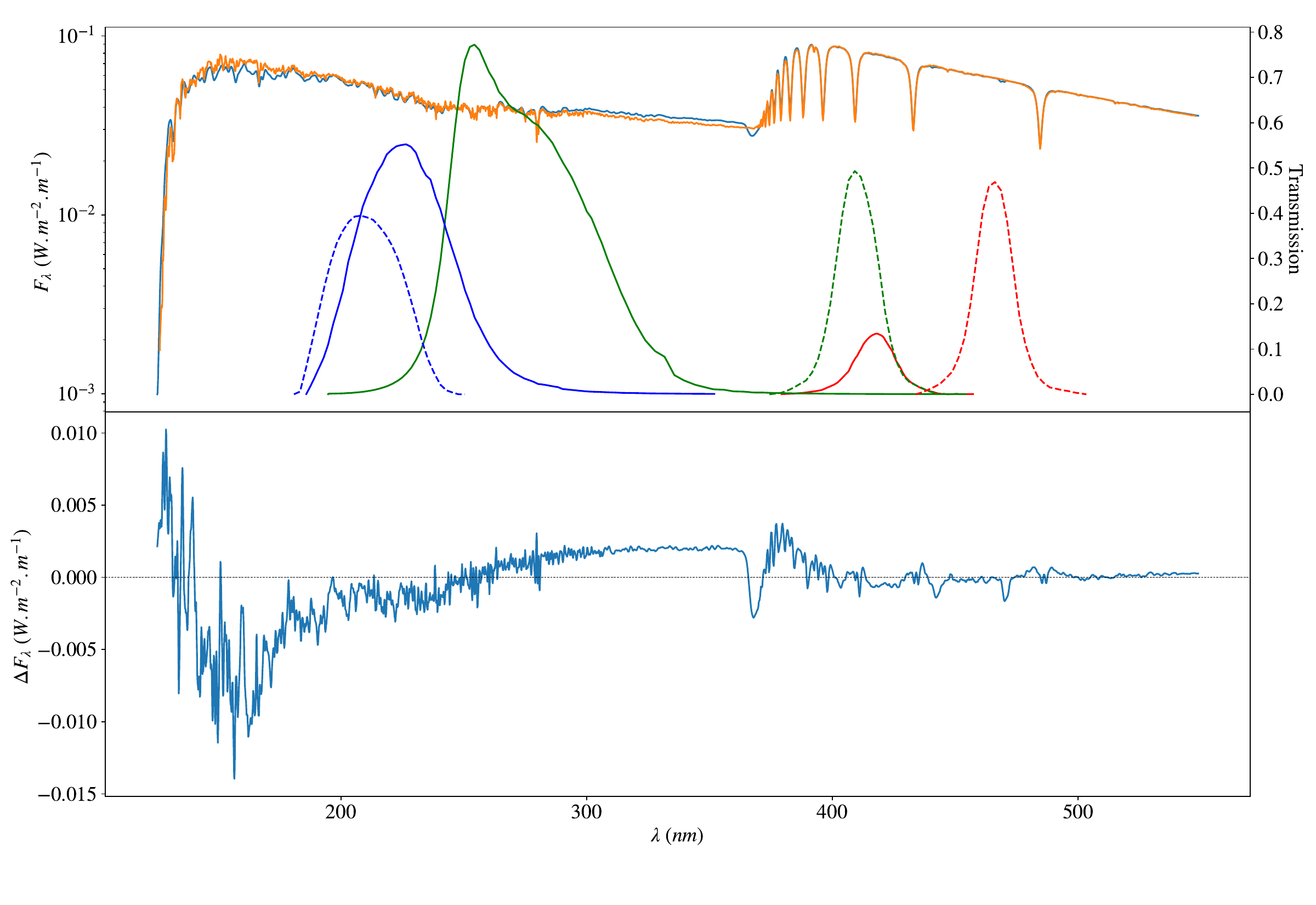} \caption{Comparison between the spectrum from CALSPEC and from our model for Vega. {\it Top panel}: Spectra of Vega from CALSPEC and from our model (orange and blue lines, respectively; left scale) and energy transmission curves of the filters used to compute the colour indices (right scale). The Str\"omgren filters are shown as  dashed lines, while the HST filters as solid lines, the colours having the same meaning as in Figs.~\ref{filters_index} and \ref{filters_index_HST}, respectively, except that the $u$ filter is replaced here by $u'$. {\it Bottom panel}: Difference between the flux of our Vega model and that of CALSPEC (black solid line).}
    \label{Vega_spec}
\end{figure*}

The next step is to compute the synthetic spectrum for this model using an inclination of $6.2^\circ$. 
We used the abundances derived from \cite{royer+14}  , which comprises a large sample of chemical elements. The abundance differences with studies that consider gravity darkening \cite[e.g.][]{yoon+10,takeda+2018} are not significant, except for carbon. Nevertheless, the global agreement between our synthetic spectrum and the observed spectrum is quite good. We show in Fig.~\ref{Vega_spec} the spectral energy distribution (SED) of our model together with its difference with the calibrated CALSPEC spectrum derived from observations \cite[e.g.][]{bohlin+14,bohlin+22}. From this synthetic spectrum we derived the three spectral observables $(T_\mathrm{IRFM}, c_1, c_2)$, which we display in Table \ref{Observable_values_Vega} along with their values derived from the CALSPEC spectrum. We note that the observed and model values of $T_\mathrm{IRFM}$ and $c_2$ agree within a 1\% error bar, while the  $c_1$ values show a mismatch of around 6\%. We trace the origin of this discrepancy to the flux difference that appears in the spectral band of the Str\"omgren $u$-filter, namely for wavelengths between 300 and 360 nm (see Fig.~\ref{Vega_spec}). Actually, this difference between the observed and synthetic spectrum has already been noted \cite[e.g.][]{castelli+kurucz94,bohlin+04}, and is explained by the difficulty faced by stellar atmosphere models to deal with the complex atomic physics needed to compute this spectral region. This mismatch between observed and synthetic $c_1$ is a serious obstacle for our inversion procedure since forcing our model to match the observed $c_1$ would necessarily lead to incorrect   stellar model parameters.

To circumvent this difficulty we chose to replace the Str\"omgren $u$-filter with a more appropriate one that does not sample this difficult region of the spectrum. Namely, we consider a new filter $u'$, which has the same shape as the $u$-filter, but is centred at a shorter wavelength $\lambda_\mathrm c$.

\begin{figure}[t!]
    \centering
    \includegraphics[width=\linewidth]{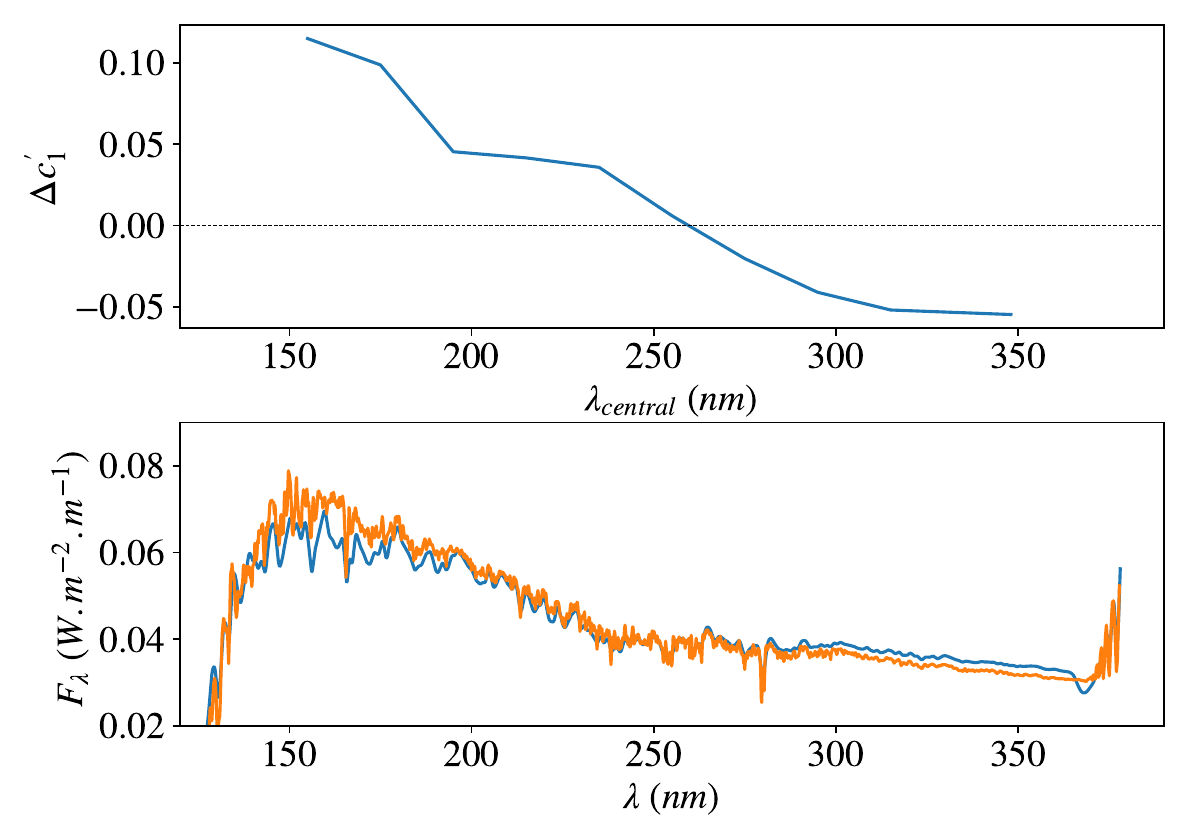} \caption{Influence of the central wavelength of the $u'$ filter on the $c_1'$ index. {\it Top panel}: Difference between the $c_1'$ index determined with our model and that of the CALSPEC spectrum as a function of the central wavelength of the $u'$ filter. {\it Bottom panel}: Spectra of Vega from CALSPEC and from our model (orange and blue lines, respectively).
 }
    \label{mv_c1}
\end{figure}

To determine the best wavelength for this, we computed the index $c'_1=(u'-v)-(v-b)$ scanning the spectrum from $\lambda_\mathrm c=170$~nm to 350~nm and computing the difference between $c'_1$ of our Vega model and that of the CALSPEC spectrum. The result is visualised in Fig.~\ref{mv_c1}, where we see that the best match occurs at $\lambda_\mathrm c=215$nm. The corresponding  $c'_1$ values are reported in the last column of Table~\ref{Observable_values_Vega}. The difference between CALSPEC and our model is now much less than a percent.

\begin{table}[t]
\begin{center}
\begin{tabular}{|c|c|}
\hline
$M$ ($M_\odot$) & 2.374 \\
\hline
$R_\mathrm{p}$ ($R_\odot$) & 2.420 \\
\hline
$R_\mathrm{e}$ ($R_\odot$) &  2.728 \\
\hline
$\epsilon$ &  0.113 \\
\hline
$L$ $(L_\odot)$ & 48.0 \\
\hline
$T_\mathrm{eff_p}$ (K) &  10070 \\
\hline
$T_\mathrm{eff_e}$ (K) &  8970\\
\hline
$\log g_\mathrm{p}$ &  4.046 \\
\hline
$\log g_\mathrm{e}$ &   3.815 \\
\hline
$v_\mathrm{eq}$ (km.s$^{-1}$) & 205 \\
\hline
$\Omega_\mathrm{e}$ ($10^{-4} \mathrm{rad.s}^{-1}$) &  1.082\\
\hline
$\Omega_\mathrm{c}$ ($10^{-4} \mathrm{rad.s}^{-1}$) &  3.067\\
\hline
\end{tabular}
\end{center}
\caption{Parameters of the ESTER Vega model of \cite{ELR13} that matches the observational constraints (interferometric and spectroscopic) derived by \cite{monnier_etal12}. For that model, $\omega=0.504$. Indices $e$ and $p$ refer to equatorial and polar values, respectively.}
\label{Vega_param}
\end{table}

\begin{table}[t!]
\centering
    \begin{tabular}{|c|c|c|c|c|}
\hline
& $T_\mathrm{IRFM}$(K) & $c_1$ & $c_2$ & $c_1'$ \\
\hline

CALSPEC     & 9662.28 & 1.154 & 4.853 & 0.6347  \\

Our model   & 9604.51 & 1.089  & 4.908 & 0.6331   \\
\hline
    \end{tabular}
\caption{$T_\mathrm{IRFM},c_1$, $c_2$, and $c_1'$ values for Vega }
    \label{Observable_values_Vega}
\end{table}

\subsection{Inversion of Vega data}

The challenge is now to use the three observed indices $T_\mathrm{IRFM}$, $c_1'$, and $c_2$ in Table~\ref{Observable_values_Vega} obtained with the CALSPEC spectrum to retrieve the fundamental parameters $M$ and $\omega$ of Vega along with the inclination of its rotation axis on the line of sight. In the present case we do not start from scratch and assume that the evolutionary state of Vega is known. Specifically, we consider a set of ESTER models computed with the metallicity and core hydrogen mass fraction of Vega. The model atmospheres and spectra were calculated accordingly. 

\begin{figure}[t!]
    \centering
    \includegraphics[width=\linewidth]{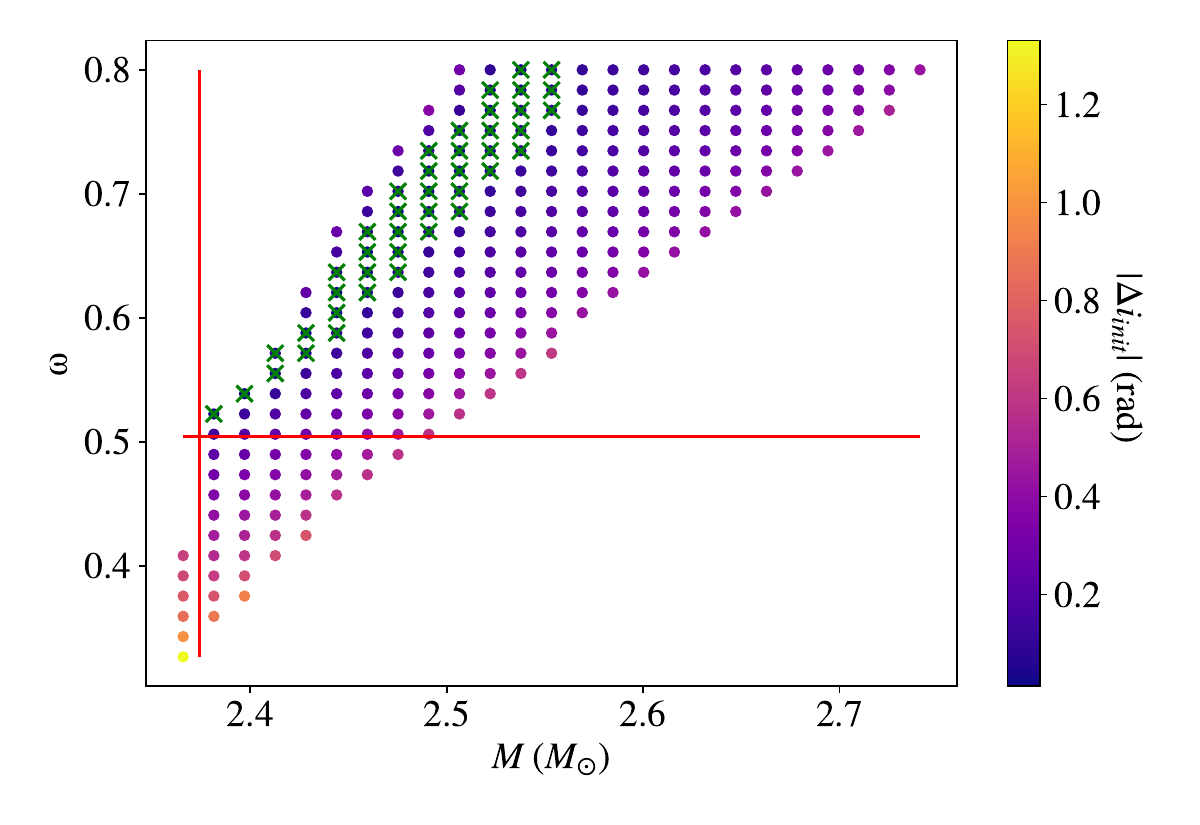} \caption{Same as Fig.~\ref{Filtrage 2}, but  for Vega.}
    \label{filtrage_vega}
\end{figure}

From the value of $T_\mathrm{IRFM}$ we deduce that the mass should be between $2~M_{\odot}$ and $3~M_{\odot}$. We thus scan this mass range to derive the analogue  of relation (\ref{empir}), but with $x_c=0.271$. From the observed CALSPEC value of $T_\mathrm{IRFM}$ we find that the approximate mass of Vega is $2.46~M_{\odot}$ with a typical  uncertainty of 15\%. This suggests  scanning a mass range of [2.1, 2.8]~$M_{\odot}$. Since we have no idea of the rotation rate, except the indication from the $V\sin i$ that the equatorial velocity is higher than 21.2~km/s, we decided to scan the whole interval of $\omega\in[0.3, 0.8]$ and to inspect the resulting inclinations. The subsequent set of initial guesses is illustrated in Fig.~\ref{filtrage_vega}. This diagram already shows that Vega must be a rapid rotator since inclination-consistent solutions (green crosses) point to $\omega$ larger than 0.5. Owing to the low value of $V\sin i$ we can already deduce that Vega is seen nearly pole-on.

\begin{figure}[t!]
    \centering
    \includegraphics[width=\linewidth]{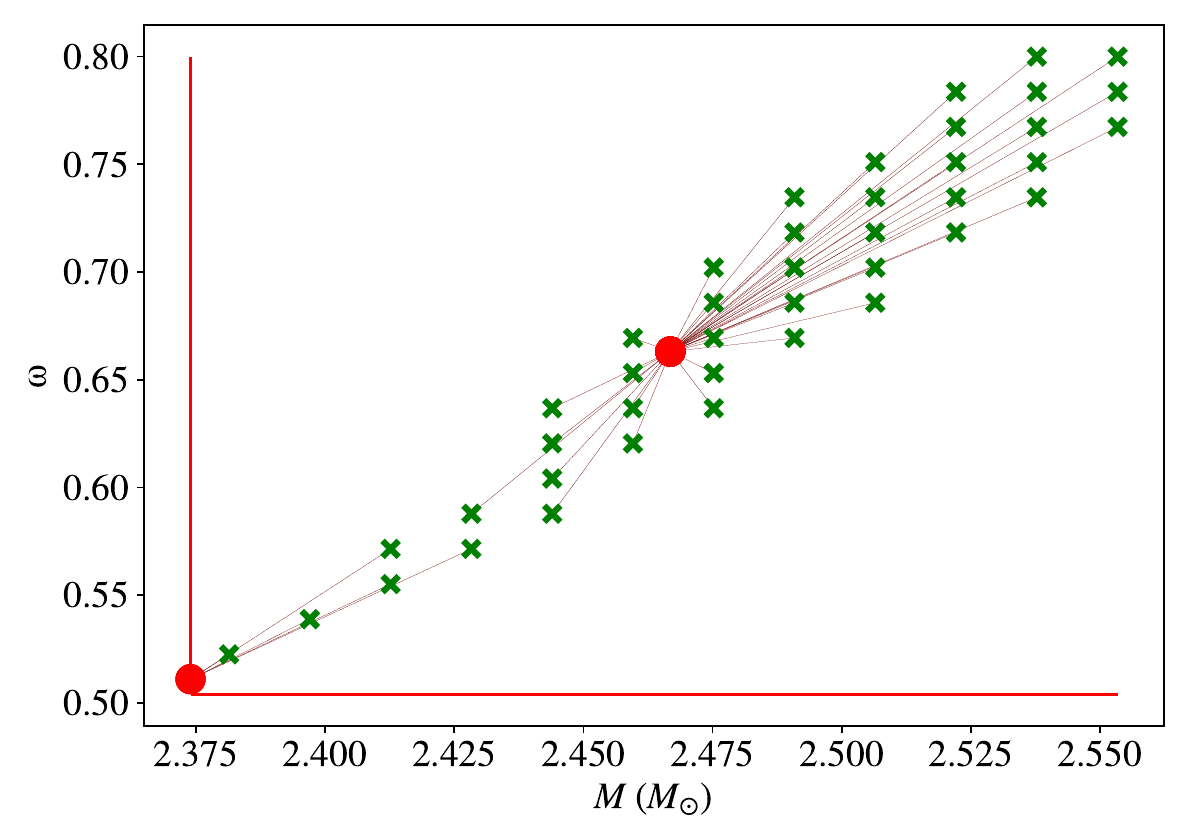} \caption{Initialisation models (green crosses) linked to their solutions (red dots) by solid lines in the $(M,\omega)$ plane. The red lines show the location of the target star.}
    \label{Solution_vega}
\end{figure}

As illustrated in Fig.~\ref{Solution_vega}, the selection steps resulted in 42 initial guesses, which converged towards two solutions: 4 to a model with $M=2.374~M_{\odot}$, $\omega=0.511$, and $i=0^\circ$, the others to a solution with $M=2.467~M_{\odot}$, $\omega=0.663$, and $i=32.1^\circ$, all with negligible uncertainties. The existence of these two solutions is due to some degeneracy in the observables. We traced back this degeneracy to the limited precision of our calculations caused by the interpolations between gridpoints. However, the accuracy of the observable determination is not infinite, which is also a source of degeneracy of the solutions. However, the solution with $M=2.467~M_{\odot}$ yields $V\sin i=137~\mathrm{km/s}$, which is easily ruled out by the observed value of $V\sin i= 21.2$~km/s \citep{takeda_etal08}. If we take the $M=2.374~M_{\odot}$ model and combine it with the observed $V\sin i$, we get $i=6^\circ$ as expected.

The example of Vega shows that our method can reveal the high rotation rate of a star, but may lead to trouble by degeneracy in some parameter range. 


\section{Discussion}
\label{section:discussion}

These results have raised  several questions. Within the assumptions we used, namely the fixed stage of evolution of our models, we can wonder about the impact of uncertainties that naturally come with measurements. For instance, if the three observables are derived with a one percent precision, we expect a precision of a few percent, but polar and equatorial orientations tend to increase the influence of uncertainties. A slow rotation ($\omega\raisebox{-.7ex}{$\stackrel{<}{\,\sim\,}$} 0.3$) producing a weak gravity darkening also amplifies errors on modelling, and this is why we did not consider such cases. Lastly, the example of Vega has shown that some degeneracy may prevent us from determining the parameters unambiguously. In this  case our method shows that Vega is a rapid rotator with $\omega\geq0.5$ and offers two possible solutions, but we cannot decide between the two. We have to resort to the value of $V\sin i$ or to the rotation period to lift the degeneracy. Scanning our grid of models and systematically testing inversions showed that a degeneracy like the previous one especially affects the mass range [2, 3.5]~$M_\odot$. The use of other spectral quantities or additional photometric indices may help circumvent such difficulties. This  option still needs to be fully explored.

The next question is obviously the determination of the evolutionary stage of the star from the spectroscopic data. From the location in the Hertzsprung--Russell diagram, giants can be separated from main sequence stars. Among the latter, as a star expands during its main sequence evolution, the radius would be the main indicator. To evaluate it, a measurement of the surface brightness and the ensuing angular diameter \citep{dibenedetto05} together with a reliable parallax may help determine the  relevant  starting models for our inversion method.

According to Fig.~\ref{Rap_IRFM_Xc1}, the $Q$ factor varies little with mass at a given age. We wanted to know  how this factor may change with age or metallicity. For this we computed $Q$ factors for models with solar metallicity at the evolutionary stage of Vega (i.e. $x_\mathrm c=0.271)$ on the one hand, and models with Vega's composition on the other (ZAMS and $x_\mathrm c=0.271$). At ZAMS, the ratio of $Q$ between solar and Vega metallicity models is very close to unity  with a dispersion close to $10^{-3}$. Hence, $Q$ depends very little on metallicity. The same seems to be true with respect to the stage of evolution when we compared the $Q$ values of ZAMS models with models at the stage of evolution of Vega, both computed with Vega's abundances. Here too, we note that the ratio is around unity and the scatter is  a few thousandths. This latter result is interesting as it allows us to use tabulated values of $Q$ to simulate the effect of rotation on isochrones of stellar clusters. Comparisons between  simulations and data would lead to new constraints on cluster ages.

\section{Conclusion}

In the present work we studied the effect of gravity darkening on three spectral quantities and examined the possibility to retrieve two fundamental parameters of a fast rotating star and the inclination of its rotation axis on the line of sight. To test this idea, we used the 2D-ESTER model of \cite{ELR13} combined with the stellar atmosphere model PHOENIX \citep{hauschildt+99} to compute synthetic spectra for a grid of models representing rapidly rotating stars of intermediate mass. Hence, our grid scans the mass range $2~M_\odot$ to $7~M_\odot$ and rotation rates from 30\% to 80\% of the critical angular velocity. Synthetic spectra were   computed for any inclination of the rotation axis on the line of sight.

With this grid of models and the associated grid of spectra, we computed the temperature derived from the infrared flux method $T_\mathrm{IRFM}$, the Str\"omgren index $c_1=(u-v)-(v-b),$ and a new index $c_2$ using HST filters in the UV domain. The  indices $c_1$ and $c_2$ are sensitive to the shape of the spectrum near the Balmer jump, while $T_\mathrm{IRFM}$ reflects the global shape of the spectrum, all three quantities being sensitive to the gravity darkening. Our goal is to derive the mass $M$, the rotation rate $\omega$, and the inclination $i$ of the rotation axis from the triplet $(T_\mathrm{IRFM}, c_1, c_2)$, restricting our computations to ZAMS models of solar metallicity in this preliminary exploration. Since our grid is rather coarse and the dependence of $(M, \omega, i)$ versus $(T_\mathrm{IRFM}, c_1, c_2)$ is non-linear, we have to resort to interpolations to solve the inversion. This introduces some numerical noise that limits the precision of the inversion, and in a few cases prevents the computation of the right solution. This drawback can be  circumvented  somewhat by a refinement of the grids, but at higher computational cost.

We first tested the method with a hare-and-hound exercise using our grid of the  ZAMS model. As the results were satisfactory (Figs.~\ref{error_maps_clamped} and \ref{intergrid}), we moved to a test with real data, considering the observed spectra of Vega from the CALSPEC database.\footnote{\tt https://www.stsci.edu/hst/instrumentation/\\
reference-data-for-calibration-and-tools/\\
astronomical-catalogs/calspec} 
After designing a 2D model of Vega that meets the data from interferometric and spectroscopic observations, we computed the synthetic spectrum of this star and derived the three observables $(T_\mathrm{IRFM}, c_1, c_2)$. We also computed these three values using the calibrated spectrum from CALSPEC and found that $c_1$ differed from the synthetic value by 6\%, while $T_\mathrm{IRFM}$ and $c_2$ agreed within a percent. The discrepancy between the observed $c_1$ and the synthetic index has been traced back to the known discrepancy between the synthetic spectrum and the observed one at wavelengths just below the Balmer jump \citep[e.g.][]{hauschildt+99,bohlin+04}. Since such a mismatch prevents our algorithm from working properly, we replaced the Str\"omgren $u$-filter, centred at 350~nm, by another filter of the same profile, but centred at 215~nm. At this  wavelength the new index $c'_1$ has the same value in both the observed and the synthetic data. To test our inversion method we computed a grid of 2D-ESTER models with masses ranging from 2 to 3 solar masses and rotation rates from 30\% to 80\% of the critical rate, at an age (defined by the core hydrogen content) coming from other studies. Browsing within this grid we could immediately conclude that Vega is a fast rotator. However, we encountered a difficulty since our algorithm produced two solutions, revealing a degeneracy in the mass range of Vega, namely between $\sim 2~M_\odot$ and $\sim 3.5~M_\odot$. In the case of Vega, this degeneracy is immediately lifted with the use of the $V\sin i$ value, which is compatible with only one of the models.

These first results are quite encouraging, but our method can still be improved. An obvious way is to reduce the interpolation errors by a finer grid in mass and rotation rates, but additional photometric indices may also be helpful. Moreover, we think that the method can be extended to age determination within the main sequence for stars of intermediate mass with a fast rotation, provided that additional information sensitive to the stellar radius is included.

The limits of our approach can be summarised as follows:
\begin{enumerate}
\item The gravity darkening should be strong enough, and it seems that this corresponds to a rotation rate above 30\% of the critical rate.

\item The model should also be precise enough; at the moment 2D-ESTER models are reliable in the mass range $1.8~M_\odot$ to $8~M_\odot$. At lower masses surface convection has an impact on the atmosphere while at higher masses, radiative mass loss also perturbs the atmosphere. Both effects are presently ignored in the standard ESTER models. In addition, the stellar atmosphere model also needs  to be closer to reality. The difficulties met at wavelengths just below the Balmer jump prevent us from using the Str\"omgren $u$-filter.

\item The third limit comes from the fact that we largely use the UV part of the spectrum, which can only be measured from space.

\item To compute the temperature with the IRFM approach, we need to know accurately the bolometric flux, which is not usually an easy task.
\end{enumerate}

The accuracy of our inversion scheme  depends on the precision of the photometry. We implicitely assumed that measurements are accurate at a one percent level. More precise observations will improve the sensitivity to gravity darkening and make the inversion easier.

The next steps in the use of gravity darkening to constrain the fundamental parameters of stars will be to first test other stars for which interferometric data exist together with precise spectrophotometric data. Interferometric measurements are indeed a way of control when the radius is left as a parameter to be determined. Stars like $\alpha$ Oph ($M\sim2.2$~$M_\odot$), $\alpha$ Leo ($M\sim4.1$~$M_\odot$), and $\zeta$ Aql ($M\sim2.5$~$M_\odot$) are all rapid rotators seen equator-on with masses appropriate for ESTER models (see \citealt{ELR13} and \citealt{howarth23+} for their mass determination). They offer the opportunity to test our algorithm with another interesting inclination.

Our method heavily relies on UV data and at the moment these data are from the HST or IUE satellite, but new space missions are planned, such as the Ultraviolet Explorer (UVEX) \citep{kulkarni+21} and the Large Ultraviolet Optical Infrared Surveyor (LUVOIR) \citep{luvoir+19}.
 Star clusters are of particular interest because their members are assumed to share the same age and the same initial composition, setting a strong constraint on these parameters. Our method can be used to determine, from simple photometry, the presence of fast rotation among stars in a cluster and its actual impact on their evolutionary stage. The HAYDN mission \citep{miglio+21} could bring valuable information about the spin axis distribution among clusters.

\bigskip
\begin{acknowledgements}
We would like to thank the referee Y. Frémat for his careful reading of the manuscript, and whose questions and comments helped to improve it. 
Most of the calculations have been performed on CALMIP
supercomputing center (Grant 2023-P20025). We also acknowledge the support of the French
Agence Nationale de la Recherche (ANR), under grant ESRR (ANR-16-CE31-0007-01).
\end{acknowledgements}

\bibliographystyle{aa}
\bibliography{Article_def}
\end{document}